\PassOptionsToPackage{dvipsnames}{xcolor}	
\documentclass[12pt,a4]{article}

\usepackage[T1]{fontenc}
\usepackage{comment}
\usepackage{lmodern}
\usepackage{setspace}
\usepackage{etoolbox}
\makeatletter
\patchcmd{\appendix}{\@Alph}{\@Roman}{}{}
\makeatother

\usepackage{amsmath}
\usepackage{amssymb}
\usepackage{amsthm}
\usepackage{mathtools}
\usepackage{mathrsfs}
\usepackage{breqn}
\usepackage{bigints}

\usepackage[margin=1 in]{geometry}
\usepackage{enumerate}
\usepackage{enumitem}
\setlist[enumerate,1]{label=(\arabic*)}
\setlist[itemize,1]{label=--}    
\usepackage{xcolor}
\usepackage{hyperref}
\usepackage{float}
\usepackage{indentfirst}
\usepackage{caption}

\usepackage{sectsty}
\sectionfont{\centering}
\subsectionfont{\centering}

\usepackage{tikz}
\usetikzlibrary{decorations.pathreplacing}
\usepackage{mathtools}
\usetikzlibrary{positioning}
\usepackage{graphicx}
\usepackage{pgfplots}
\pgfplotsset{compat=1.18} 
\definecolor{dodgerblue}{rgb}{0.1211,0.5664,1}
\definecolor{golden}{RGB}{255, 206, 84}
\definecolor{chocolate}{rgb}{1,0.5,0.14}
\definecolor{burgundy}{rgb}{0.78125, 0.1, 0.246}
\definecolor{budgreen}{rgb}{0.53125, 0.7226, 0.3125}
\definecolor{almond}{rgb}{0.985, 0.922, 0.785}

\newcommand{\bb}{\mathbb}
\newcommand{\scr}{\mathscr}
\newcommand{\U}{\bigcup}
\newcommand{\A}{\bigcap}
\newcommand{\und}{\underline}
\newcommand{\lims}{\lim\limits}

\newcommand{\mcal}{\mathcal}

\newcommand{\supp}{\text{supp}}

\renewcommand{\epsilon}{\varepsilon}

\newcommand{\blue}[1]{\color{blue}#1 \color{black}}

\newtheorem{theorem}{Theorem}
\newtheorem{lemma}{Lemma}

\newtheorem{proposition}{Proposition}
\newtheorem{corollary}{Corollary}
\newtheorem{assumption}{Assumption}

\newtheorem*{statement*}{Result}

\theoremstyle{definition}
\newtheorem{definition}{Definition}

\DeclareMathOperator*{\argmax}{\arg\!\max}

\DeclareTextFontCommand{\emph}{\slshape}

\usepackage{natbib}
\nocite{*}

\title{Paying and Persuading}
\author{Daniel Luo\thanks{Massachusetts Institute of Technology, daniel57@mit.edu. I am especially grateful to Drew Fudenberg, Stephen Morris, and Alex Wolitzky for invaluable guidance. I am thankful to Zihao Li for discussions at the beginning of the project, and Ian Ball, Abhijit Banerjee, Alessandro Bonatti, Yifan Dai, Laura Doval, Eric Gao, Marina Halac, Navin Kartik, Andrew Koh, Anton Kolotilin, Andrew Komo, Elliot Lipnowski, Delong Meng, Meg Meyer, Ellen Muir, Harry Pei, Ryo Shirakawa, Eric Tang, Frank Yang, Kai Hao Yang, the Fall 2025 3B Conference, MIT Theory Lunch, MIT Organizational Economics Lunch, Yale Theory Breakfast, and Stonybrook 2025 for helpful suggestions and comments. Coarse.ink and Refine.ink were used to check this manuscript for clarity and correctness. I acknowledge financial support from the NSF Graduate Research Fellowship. All errors are mine alone.}}
\date{\today}

\begin{document}
\maketitle

\begin{abstract}
I study dynamic contracting where Sender privately observes a Markovian state and seeks to motivate Receiver, who acts. 
Sender provides incentives in two ways: payments, which alter payoffs ex-post, and (Bayesian) persuasion, which shapes Receiver interim beliefs about payoffs. 
For all stage game payoffs, discount rates, and Markov transition rules, \emph{transfers are a last resort}---there is an optimal contract where payments occur only after Sender commits to reveal the state at every continuation history. 
In an example, the optimal contract is a \emph{loyalty program}: Sender chooses the static optimal information structure until a random promotion time, after which Sender reveals the state and pays Receiver. 
\end{abstract}


\newpage

\section{Introduction}
\onehalfspacing

Consider the problem faced by a principal who wishes to induce an agent to take an ex-ante undesirable action. Standard economic theory suggests two ways to motivate such an agent: directly compensate them for taking such an action with a monetary transfer (paying), or manipulate their information environment so that the action becomes interim desirable (persuading). In settings where only one such tool is available to the principal\footnote{In moral hazard or incomplete contracting for the former, or information design for the latter.},  much is known about the optimal way to pay \textit{or} persuade, and economists have derived various solutions in increasingly sophisticated models of both, separately. 

Yet little is known about the optimal way to pay \textit{and} persuade---to jointly compensate an agent for taking certain actions while also changing the expected value of those actions by manipulating the agent's informational environment. That such joint incentive problems are understudied is, in part, historical. For example, in models of moral hazard, the agent is the one traditionally endowed with the informational advantage, not the principal, and so there is no room for information design. Likewise, in models of information design, such as the judge-prosecutor example of \cite{KamenicaGentzkow11} or FDA clinical trials for drug discovery, it is unnatural to suppose Sender also pays Receiver. Consequently, the intersection of these two tools is not very well understood.

Despite this prima facie dichotomy between paying and persuading, there are new economic settings where a principal jointly controls information and monetary transfers. For example, in addition to varying payments, Uber strategically varies the amount of information drivers can see about recommended trips in order to motivate their drivers to accept rides that they may find ex-post undesirable.\footnote{Uber writes 
``The type of approximate pickup and dropoff location shown to drivers varies based on address structures in your market, local regulations, and driver loyalty programs.'' (\cite{uber2025privacy}).} 
Advertisers both run advertisements and email private offers that discount the good in order to incentivize purchases. 
Managers who privately observe the value or difficulty of a project may both strategically provide feedback to the worker and directly vary their compensation scheme in order to convince them to work when they otherwise might prefer to shirk. 

Taking seriously the interaction between paying and persuading raises several new questions.
What is the optimal way to combine information and transfers in a one-shot or a repeated interaction?
When the principal and agent interact repeatedly, what can be said about the trade-off between using future information and future payments to motivate the agent? 
Since information generation is free, is it ever profitable to dynamically withhold information as a way to incentivize Receiver? 
What is the value of repeated interaction when the principal has access to both tools, and how does it compare to the one-shot game? 

This paper studies dynamic contracting where Sender can both flexibly manipulate information and offer payments to Receiver. Formally, I consider a model where a principal (Sender) commits to (1) provide information about a Markovian state and (2) a contingent transfer rule prior to an agent (Receiver) choosing a perfectly monitored action in each period. 
Sender can commit to the entire path of transfers and information, conditioning payments and persuasion today based on anything in the past history.\footnote{For example, Uber writes the algorithm for, and publicizes, their dynamic reward structure simultaneously, committing to rewarding drivers in the future for past performance.}
Information and transfers thus have an intertemporal component: while payments can occur immediately to change behavior today, they can also be leveraged in the future as a dynamic reward. 
Thus, when seeking to motivate Receiver, Sender must balance the use of information and transfers both within each period and across time.  

My main result is a striking backloading theorem that characterizes the tradeoff between optimal information and transfers in Sender's dynamic contracting problem. 
Formally, Theorem \ref{t: transfers as last resort} constructs an optimal contract where, if transfers are ever used, Sender has already committed to fully reveal the state at all continuation histories, i.e. committed to relinquishing their informational advantage at every point in the future. 
This characterization applies for any Sender or Receiver stage-game payoffs, discount rate, and law governing the Markov chain, and thus speaks generally to the way information and transfers trade off against one another in dynamic contracts. Theorem \ref{t: transfers as last resort} can be interpreted as a sequencing argument: jointly paying and persuading Receiver tomorrow must always be weakly as efficient as only paying Receiver today, and hence there is an optimal contract where payments are made on the path of play only once information is exhausted as a dynamic incentive instrument. 

The proof hinges on a \textit{pullback} argument (Lemma \ref{l: pullback lemma}), which establishes that the full-information experiment lies on the Pareto frontier.\footnote{The Pareto frontier is the graph of Sender's value function $V(u)$, i.e. their value from optimally providing utility $u$ to Receiver.} 
This result depends on a careful interaction between information and transfers. In particular, it must show that transfers alone are not a cheaper way to provide incentives when compared to fully revealing the state. A-priori, one might worry that disclosure might be inefficient because Receiver would respond to new information in ways particularly unfavorable to Sender.
Lemma \ref{l: pullback lemma} ameliorates this concern by showing that a joint payment-and-persuasion scheme must be at least as efficient as using payments alone. The key idea is simple: Sender can, after revealing the state, always pay Receiver to take any action that Sender wants at a constant efficiency cost. Thus additional information revelation cannot impose a higher utility cost than simply paying for the action Receiver would have taken in the absence of additional information. Consequently, combining future payments with persuasion is weakly more efficient than making an immediate payment, creating incentives to backload transfers. This illustrates how transfers fundamentally change the opportunity cost of information revelation in dynamic persuasion.

Following Theorem \ref{t: transfers as last resort}, I characterize several properties of actions and beliefs that hold uniformly under all optimal contracts for i.i.d. states. First, Propositions \ref{p: feasibly optimal actions} and \ref{p: strict optimality} identify testable conditions on the sets of actions and beliefs consistent with any optimal contract, regardless of whether transfers are fully backloaded.\footnote{Recall Theorem \ref{t: transfers as last resort} establishes one such contract where transfers are backloaded: Propositions \ref{p: feasibly optimal actions} and \ref{p: strict optimality} identify properties that hold uniformly across all optimal contracts. These restrictions are based on the ratio of Sender to Receiver payoffs when switching an action from $a$ to $a'$ at a belief $\mu$, an object I call the \emph{effectiveness ratio}.} 
Second, Propositions \ref{p: weak ray regularity} and \ref{p: wrr if static k-cavification} identify conditions where \textit{all} optimal contracts backload transfers, strengthening the conclusion of Theorem \ref{t: transfers as last resort} under additional regularity conditions. These conditions ensure information and transfers are not always perfect substitutes, which is sufficient to guarantee that informational incentives always take strict precedence over transfers as a motivating tool. Finally, Proposition \ref{p: always last resort} shows eveyr optimal contract backloads transfers whenever transfers are sufficiently expensive. 

Next, I compute the value of paying and persuading in several specialized contexts. First, I consider the static joint design problem. Here, I show the value of joint information and transfers takes the form of an augmented concavification---which I call the $\mcal K$-cavification (Proposition \ref{p: k-cavification}). The $\mcal K$-cavification requires Sender to compute their optimal transfer only at each of finitely many extremal beliefs---those where Receiver is indifferent between maximally many actions---and then linearly interpolates between the value at each of these beliefs to obtain the value of joint payments and persuasion. This iterative procedure dramatically simplifies the complexity of finding the value of jointly paying and persuading and gives simple conditions under which dynamic incentives are strictly optimal in the general model. 

Second, I bound the value of dynamically paying and persuading as players become very patient when the state transition process is ergodic. Sender's value from the optimal dynamic contract must be bounded from below by their best payoff from all static couplings over states and actions that satisfy a Receiver individual rationality constraint evaluated at the ergodic distribution of the transition process. Such a characterization implies simple conditions under which Sender's dynamic value strictly outperforms their one-shot persuasion value in the static game, and thus implies conditions where dynamic incentivization benefits Sender. It also allows me to comment on the literature on Markovian persuasion, and show that when Receiver is not myopic, the canonical upper bound established in the literature instead becomes a lower bound (Proposition \ref{p: persuasion upper bound}). 

Finally, I explicitly characterize the optimal contract in a rideshare-inspired example. Uber (Sender) repeatedly incentivizes a driver (Receiver) to accept multiple rides in each period.\footnote{The stage game nests the canonical binary-state, binary-action example from the persuasion literature as a special case.}
When ride values are independently drawn across periods and utilities are additively separable, the optimal contract takes the form of a tiered \emph{loyalty program}:
In each tier, Uber chooses the static optimal information structure until a random promotion time, after which they reveal some states and simply pay the Receiver. 
The optimal contract therefore features distinct, ranked phases; in each phase, Uber tracks acceptance behavior for some number of periods and strategically manipulates the driver's informational environment. As the relationship matures, it moves up in rank, and Uber begins to reveal more information about ride quality. Moreover, they allow the driver to reject undesirable rides without penalty, relying directly on payments for incentives. This characterization is reminiscent of stylized, publically available facts regarding Uber's loyalty program for their drivers. 

The remainder of this paper is organized as follows. I next review the related literature. Section 2 introduces the general model. 
Section 3 states and proves my main backloading result. Section 4 characterizes properties that hold uniformly across all optimal contracts. Section 5 gives related results characterizing the value of persuasion when discounting approaches either $0$ or $1$. Section 6 specializes the payoff and information structure and characterizes explicitly the optimal contract. Section 7 concludes. Appendix A contains relevant omitted proofs. The supplemental appendix collects miscellaneous technical details and related extensions and examples.

\subsection{Related Literature}

I relate to several distinct strands of literature. First is the literature on dynamic contracting, as in \cite{thomas1988self} and \cite{thomas1990income}, who study the time structure of optimal transfers with private information about income and outside options.
Most closely related to my paper is \cite{ray2003time}, who characterizes in a general model the time structure of optimal transfers; in contrast, I allow the principal to control information about an unknown state and characterize the time structure of optimal transfers \emph{and persuasion.} Augmenting the model with persuasion is substantial; while the technical motivation undergirding our papers is similar, the resulting economic forces and mathematical structure that characterize optimal contracts are quite distinct.\footnote{See the discussion succeeding Theorem \ref{t: transfers as last resort} for details.} 

Methodologically, I pull from the literature that characterizes the optimality of dynamic contracts by tracking the geometry of the Pareto frontier for incentives, an approach initiated by \cite{spear1987repeated}. Subsequently, it has been used in dynamic moral hazard problems by \cite{LipnowskiRamos2020} and \cite{guo2020dynamic}, relational contracting by \cite{LiMatouschek2013}, political economy by \cite{Yared2010}, \cite{PadroIMiquelYared2012}, and \cite{AcharyaLipnowskiRamos2024},
and dynamic verification by \cite{LiLibgober2024}. I contribute to this literature by using the properties of the Pareto frontier to characterize the tradeoff between two distinct tools Sender can access, and apply these methods to dynamic information design. 


As Sender in my model controls the flow of information to Receiver and slowly ``unwinds'' their informational advantage, my model also has some qualitative features reminiscent of apprenticeship models (see \cite{FudenbergRayo2019} and \cite{FudenbergGeorgiadisRayo2021}). Though the specific ways information is modeled are distinct, information is inefficiently withheld at the beginning of both our models and then gradually provided to Receiver as a motivating tool. Eventually, this advantage is relinquished and the principal begins to pay the agent. Thus I provide another parsimonious way to model apprenticeship dynamics when information is not literally a stock or irreversible but simply provided via communication.

Importantly, I suppose the principal can flexibly design the information structure, in the spirit of the Bayesian persuasion literature started by \cite{rayo2010optimal} and \cite{KamenicaGentzkow11}. A rich literature has since pushed the limits of static persuasion; see \cite{BergemannMorris2019} for a survey. 
Yet work on dynamic persuasion is sparser. \cite{ely2017beeps}, \cite{renault2017optimal} and \cite{LehrerShaiderman2025} assume a myopic Receiver and characterize when greedy policies are optimal, while \cite{ball2023dynamic} solves the full dynamic persuasion problem but only for a linear-quadratic game with a Brownian state. Further afield, \cite{ElyFrankelKamenica2015} study optimal information provision when Receiver has preferences for noninstrumental information.  
\cite{DovalEly2020} and \cite{MakrisRenou2023} characterize the set of all possible outcomes attainable in a dynamic information design problem, but leave open the question of optimal information design. 
In the absence of commitment, \cite{GolosovSkretaTsyvinskiWilson2014} and \cite{RenaultSolanVieille13} study repeated cheap talk games and characterize when full revelation is possible and the payoff set of the patient game, respectively. 

The literature on persuasion with transfers is similarly sparse. \cite{li2017model} solves an example where Sender can pay and persuade in a static setting with binary actions, binary states, and state-independent payoffs. Further afield, \cite{ravid2022learning}, \cite{terstiege2022competitive}, and \cite{YamashitaZhu2023} study the joint design of information and transfers in auctions and screening, assuming the individual who persuades (Seller) is distinct from the one who pays (Buyer). 
In the one-shot context, \cite{DovalSkreta2022} fully characterize the optimal joint paying-and-persuading model in a mechanism design context by recasting limited commitment on the designer's part as a Bayes plausibility constraint. 
Finally, in the dynamic context \cite{HornerSkrzypacz2016} consider a binary-state game where the persuader is paid transfers for revealing information, and show that multi-stage communication is always beneficial to the high-type agent. Relatedly, \cite{MadsenWilliamsSkrzypacz2026} study a static, reduced-form model where a principal trades off between quasilinear transfers and convex amenity tools, interpreting the convex cost as continuation utility in a dynamic interaction reminiscent of my setting. 
To the best of my knowledge, this paper is the first to embed transfers into a dynamic model of persuasion, and the first to establish qualitative properties of the optimal mechanism. I also provide a new characterization of the value of dynamic persuasion with payments, which nest the value of dynamic persuasion alone (by taking the cost of transfers to be arbitrarily large). 

Also related is the literature on information design in optimal stopping problems (\cite{ely_szydlowski2020}, \cite{OrlovSkrzypaczZryumov2020}, \cite{koh2022attention} and \cite{koh2024persuasion}), which characterizes the optimal mechanism when Receiver takes an action only once and the state is perfectly persistent. Importantly, flexible endogenous transfers are absent in this literature. Consequently the economic forces driving our models are quite different. 
Particularly related is \cite{ely_szydlowski2020}, who study optimal dynamic information design in a continuous time, work-or-shirk model when the transfer scheme is fixed and paid out only at the end; I show that when transfers can be endogenously combined with information in any flexible fashion, payments are paid out only after the initial (pathwise) period of pure information design. 

I also draw from the literature on information design in moral hazard (\cite{ely2022optimal} and \cite{ely_rayo_2024feedback}). These papers allow for transfers, but must assume a perfectly persistent binary state as they seek to characterize the optimal mechanism explicitly. In the case without moral hazard but where the principal has access both instruments, I show the optimal contract has a \emph{loyalty contract} structure, in contrast to their results. 

My geometric characterization of static persuasion with transfers (Proposition \ref{p: k-cavification}) contributes to the literature on the geometry of communication. I build on work that geometrically characterizes the value of communication in persuasion (\cite{KamenicaGentzkow11}), cheap talk with transparent motives (\cite{lipnowski2020cheap}), constrained persuasion (\cite{DovalSkreta2023}), and flexible information acquisition (\cite{Barros25}). 

Finally, I relate to a newer literature on persuasion in repeated games, often with monetary transfers. \cite{OrtnerSugayaWolitzky2024} and \cite{SugayaWolitzky2025} study mediated games where a designer wishes to implement some optimal outcome among players who repeatedly play a pricing game (i.e. a first price auction or a Bertrand competition market game). \cite{KolotilinLi2021} study a linear cheap-talk game and show Sender attains their best monotone contract when transfers are voluntary (i.e. relational). 
In contrast, I analyze a model where the person who persuades is also the one who pays, and give a backloading result when the principal can commit to the entire path of transfers.

\section{Model}
\subsection{The Stage Game}
There are two players, Sender (S) and Receiver (R). Sender privately observes a state $\theta$ drawn from distribution $\mu_0 \sim \Delta(\Theta)$; Receiver takes action $a \in A$. $A$ and $\Theta$ are both finite.\footnote{Except for the last step of the proof of Theorem \ref{t: transfers as last resort} and Proposition \ref{p: k-cavification}, both $A$ and $\Theta$ need only be compact sets, modulo the appropriate measure-theoretic generalization. However, to invoke the strong topology (see \blue{\ref{Appendix B}}) and compute extremal beliefs, respectively, $A$ and $\Theta$ must be finite.}

Sender can design a contract to motivate Receiver in one of two ways. First, they can transmit information, modeled as a choice of signal structure $\mcal S: \Theta \to \Delta(S)$ from states into signals, where the signal space $S$ is a subset of some standard Borel space. 
Second, they can commit to a transfer rule, $t: \Theta \times S \times A \to [0, C]$ which specifies a nonnegative\footnote{In the absence of a nonnegativity restriction, i.e. a limited liability constraint, Sender can obtain first best by ``informationally'' selling the firm to the agent---offering full information for a payment equal to their value of information, then paying them to take the surplus maximizing action.} bounded\footnote{I allow $C$ to be arbitrarily large, but finite for technical reasons; in the dynamic model, this ensures transfers satisfy a standard transversality condition that is crucial for compactness of the contract space.} payment for each pair of states, signals, and actions. Let $\mcal T$ denote the set of all transfer rules.
Having seen $(\mcal S, t)$ and realized signal $s \in S$, the agent forms beliefs and takes an action $a$ maximizing their expected transfer-augmented utility 
\[ a \in \argmax_{\tilde a \in A}\bb{E}_{\theta | s}[u(\tilde a, \theta) + t(\theta, s, \tilde a)]  \]
where $u: A \times \Theta \to \bb{R}$ is Receiver's bernoulli payoff. Say $\bar a: S \times \mcal T \to \Delta(A)$ is a \emph{best response} to $(\mcal S, t)$ if $\bar a(\cdot)$ always maximizes Receiver's expected transfer-augmented utility.

Let $v: A \times \Theta \to \bb{R}$ be Sender's bernoulli payoff. 
Following the approach of \cite{KamenicaGentzkow11} (modified for transfers), I represent a tuple $(S, t, \bar a(\cdot))$ where $\bar a(\cdot)$ best responds to $(\mcal S, t)$ as a tuple $(\rho, t, \bar a(\cdot))$ where $\rho \in \Delta_{\mu_0}(\Delta(\Theta))$ is a $\mu_0$-Bayes plausible distribution of posterior beliefs and 
\[ \bar a(\mu, t) \in \argmax_{a \in A} \bb{E}_\mu[u(a, \theta) + t(\theta, s^{-1}(\mu, \bar a), a)]. \]
Here, $s^{-1}(\mu, \bar a)$ is the signal that induces belief $\mu$ and action $\bar a(\cdot)$. 

For any transfer rule $t$, suppose Sender incurs a per-unit cost $k$ to pay Receiver $1$ unit of payoff, so that for realized state, signal, and action, $(\theta, s, a)$, Sender's payoff is given by 
\[ v(a, \theta) - k t(\theta, s, a). \]
As $k \to 0$, Sender can attain first best. If $k = 1$, this is the familiar quasilinear transferable utility case. As $k \to \infty$, transfers become arbitrarily expensive and Sender reverts to their no-transfers persuasion baseline. 
Consequently, my model can interpolate between (1) Sender first best (2) transferable utility, and (3) the persuasion benchmark.\footnote{Allowing for the efficiency of transfers to vary in $k$ will explicitly clarify the incentive role of transfers in motivating the agent, instead of complicating the exposition.}

The timing of the stage game is as follows. 
\begin{enumerate}
    \item Sender and Receiver realize a common prior $\mu_0$ over today's states and actions. 
    \item Sender commits to a signal structure $\mcal S$ and transfer rule $t$. 
    \item Nature draws $\theta \sim \mu_0$, and then signal $S$ is realized according to lottery $\mcal S(\theta)$. 
    \item Having seen $(\mcal S, t)$ and a specific realization $s$, the agent takes an action $a$. 
    \item Ex-post payoffs $u(a, \theta) + t(\theta, s, a)$ and $v(a, \theta) - k t(\theta, s, a)$ are realized. 
\end{enumerate}

An immediate simplification allows the transfers to be written only as a function of $(s, a)$, integrating out the state. This is because incentive compatibility of any recommended action depends only on the expected transfer given any signal, not the specific state-dependent ex-post payment. Say tuples $(\mcal S, t, a)$ and $(\mcal S', t', a')$ are \emph{outcome equivalent} if they induce the same expected on-path transfers and joint distribution over beliefs and actions. 

\begin{lemma}
\label{l: transfers not depend on state}
    Fix $(\mcal S, t, \bar a)$. There exists outcome equivalent $(\mcal S', t', \bar a)$ where $t'$ is constant in $\theta$.
\end{lemma}
\begin{proof}
    Define $t'(s, a) = \bb{E}_{\theta | s}[t(\theta, s, a)]$. Clearly, this induces the same expected transfers and joint distribution of beliefs and actions, maintaining incentive compatibility. 
\end{proof}

A straightforward modification extends the lemma into the dynamic game described in the next subsection. I omit the dynamic generalization for expositional succinctness. 
In principle, a revelation principle also holds (see \cite{KamenicaGentzkow11}) where it is without loss of generality for $S = A$. 
However, it turns out that it will be useful in the dynamic case for the signal space to be $S = \Delta(\Theta) \times A$, i.e. for Sender to both specify a posterior belief (a la the belief-based approach), and a recommended action (a la the obedience approach).

\subsection{Dynamics}
Suppose the stage game is played over many periods, $\xi = 0, 1, 2, \dots$ between Sender and Receiver, both of whom discount the future at a common rate $\delta \in (0, 1)$. Moreover, suppose Sender can commit to an entire history-contingent plan of action at the start of the game. Call this plan Sender's \emph{contract}.

The state, which in period $0$ is drawn according to some prior $\mu_0$, thereafter is drawn according to some arbitrary Markov chain represented by the linear operator $M: \Delta(\Theta) \to \Delta(\Theta)$. In the case where $M$ is irreducible and aperiodic, let $\mu^\infty$ be the resulting unique ergodic distribution of the stochastic process induced by $M$. 

Receiver's past actions are perfectly observed by Sender, but past states are never observed.\footnote{Unobservability of past states does not affect the conclusion of Theorem \ref{t: transfers as last resort} in the i.i.d. case. Relaxing it substantially complicates the remainder of the analysis otherwise, so I maintain it for simplicity.} Thus, a time-$\xi$ history is a sequence $h^\xi = \{s_\zeta, t_\zeta, a_\zeta\}_{\zeta = 0}^{\xi - 1}$ of past signals, transfers, and actions.\footnote{I suppose players do not observe their payoffs, so that the state is not revealed at the end of each period. When the state is drawn i.i.d., as in the leading example, this assumption is without loss of generality.} However, because transfers at time $\xi$ are measurable in history $h^{\xi - 1}$ and the time-$\xi$ signals and actions and Sender has commitment, it is without loss of generality to set histories to be sequences $\{s_\zeta, a_\zeta\}_{\zeta = 0}^{\xi - 1}$  of signals and actions only. Let $H^\xi = (S \times A)^\xi$ be the set of all time-$\xi$ histories and $H = \U_{\xi \in \bb{N}} H^\xi$ the set of all histories, which I identify with $H^\infty = (S \times A)^\infty$, the set of infinite histories in the standard way (see \cite{mertens2015repeated}). 

A contract is a tuple $\sigma = (\sigma^S, \sigma^R)$ of functions $\sigma^S: H \to \Delta(S)^\Theta \times \mcal T$ and $\sigma^R: H \to \Delta(A)^{S \times \mcal T}$ specifying stage-game strategies for Sender and Receiver at each history. 
Let $\Sigma = (\Sigma^S, \Sigma^R)$ be the set of all contracts. Endow $\Sigma$ with the strong topology (see \blue{\ref{Appendix B}}). 
For any contract $\sigma$, let $\bb{P}^\sigma \in \Delta(H)$ be the distribution over histories it induces and $\bb{Q}^\sigma \in (\Delta(\Theta \times A))^\infty$ be the joint distribution over outcomes it induces. Let $\mu(h^\xi; \sigma)$ be Receiver's time-$\xi$ prior belief about the state at history $h^\xi$ given contract $\sigma$, before any information is disclosed at time $\xi$. In particular, this implies the distribution of beliefs induced by $\sigma^S$ at $h^\xi$ must be $\mu(h^\xi; \sigma)$-Bayes plausible. 

Let $\sigma(\cdot | h^\xi) = (\sigma^S(\cdot | h^\xi), \sigma^R(\cdot | h^\xi))$ denote the continuation contracts at $h^\xi$ and $\sigma(h^\xi) = (\sigma^S(h^\xi), \sigma^R(h^\xi))$ the profile played at $h^\xi$. 
Given some $\sigma$ and $\bb{P}^\sigma$, say $\sigma$ is \emph{obedient} if at every $h^\xi \in \supp(\bb{P}^\sigma)$, Receiver's continuation contract $\sigma^R(\cdot | h^\xi)$ is optimal among the set of all possible continuation contracts: 
    \[ \sigma^R(\cdot | h^\xi) \in \argmax_{\sigma^{R'}(\cdot | h^\xi) \in \Sigma^R(\cdot | h^\xi)} \bb{E}_{\bb{P}^{(\sigma^S(\cdot | h^\xi), \sigma^{R'})}}\left[ \mcal U(\sigma^S(\cdot | h^\xi), \sigma^{R'}) \right] \]
where $\mcal U(\sigma^S(\cdot | h^\xi), \sigma^R(\cdot | h^\xi))$ is Receiver's continuation payoff given contracts at $h^\xi$. Note it is without loss of generality to consider contracts where off-path deviations are punished with no information and zero transfers forevermore. 
A contract $\sigma$ is \emph{optimal} if it is obedient and attains the maximum of Sender's payoff among all obedient contracts. 
Let $\mcal V^*(\mu_0, \delta)$ be Sender's payoff at optimum when the prior is $\mu_0$ and players discount the future at $\delta$.

The following lemma will be useful, which provides a dynamic version of the standard revelation/obfuscation principle for this setting (albeit with an expanded signal space). 

\begin{lemma}
    \label{l: dynamic revelation principle}
    Let $\sigma$ be obedient. There exists obedient $\sigma^*$ with $S = \Delta(\Theta) \times A$ such that for all $h^\xi \in \supp(\bb{P}^{\sigma^*})$, $\mu(h^\xi, \sigma^* | (\mu, a)) = \mu$, $\sigma^{*R}(h^\xi)(\mu, a) = a$, and $t((\mu, a), a'; \sigma^*) \neq 0 \implies a = a'$ such that $\bb{Q}^\sigma = \bb{Q}^{\sigma^*}$. 
\end{lemma}
The proof is a standard relabeling argument, done in \blue{\ref{l: dynamic revelation principle proof}}
Lemma \ref{l: dynamic revelation principle} implies that for each obedient contract there is an outcome equivalent (and hence payoff equivalent) contract $\sigma^*$ which is direct in beliefs and actions: it tells Receiver (directly) what belief to hold, and then recommends an incentive compatible action. Moreover, it further simplifies the space of transfers: Sender need only pay Receiver if Receiver follows the recommended action. 
This implies the set of transfers needed to be specified simplifies from one for each pair of actions only to one for each action recommendation, a much lower dimensional space. 

Why is it that Lemma \ref{l: dynamic revelation principle} does not restrict to direct action recommendations, which are sufficient in the static case? While it is also sufficient in the dynamic case to restrict directly to action recommendations only (randomizing the belief implicitly), this obfuscates policies where Sender induces multiple actions at the same posterior belief (or multiple beliefs at the same action). While it is without loss of generality to consider random actions only, doing so will introduce notational complexity in the proof of the main result (Theorem \ref{t: transfers as last resort}) so I work with this expanded signal space instead throughout the paper. 

Note Lemma \ref{l: dynamic revelation principle} implies it is equivalent to recast the problem into finding Bayes plausible distributions of beliefs $\rho \in \Delta(\Delta(\Theta))$, transfer rules $t: \Delta(\Theta) \times A \to \bb{R}_+$, and recommended (random) actions $\alpha: \Delta(\Theta) \to \Delta(A)$ such that (1) Receiver finds it incentive compatible to take all actions $a \in \supp(\alpha(\mu))$, and (2) beliefs are Bayes plausible, $\bb{E}_\rho[\mu] = \mu_0$. 
Paired with a future utility continuation promise $u': \Delta(\Theta) \times A \to \bb{R}$, this implies the following recursive formulation of the optimal contract:
\begin{align*}
\label{equation_1}
V(\bar u, \mu_0, \delta) = \max_{\{\rho, t, u', \alpha\}} 
   \Big\{ \bb{E}_\rho\Big[\bb{E}_{\mu, \alpha}[(1 - \delta)\,[v(a, \theta) & - k t(\mu, a)] 
   + \delta V(u'(\mu, a), M\mu, \delta)]\Big] \Big\}
\tag{FE} \\[1em]
\text{s.t.}\quad 
\bb{E}_{\mu}[(1 - \delta) [u(a, \theta) + t(\mu, a)] + \delta u'(\mu, a)] 
&\geq \bb{E}_\mu[(1 - \delta) u(a', \theta) + \delta \und U(M\mu)]
\tag{IC} \\
&\hspace{0.5em}\forall \mu \in \supp(\rho),\; a \in \supp(\alpha(\mu)),\; a' \in A \\[1em]
\bb{E}_\rho[\bb{E}_{\mu, \alpha}[(1 - \delta)[u(a, \theta) + t(\mu, a)] + \delta u'(\mu, a)]] 
&\geq \bar u 
\tag{PK} \\[1em]
\bb{E}_\rho[\mu] 
&= \mu_0 
\tag{BP} \\[1em]
\max\{(1 - \delta) t(\mu, a),\, |u'(\mu, a)|\} 
\in [0, C] \quad \forall  & \mu \in \supp(\rho), a \in \supp(\alpha(\mu)), 
\tag{BD}
\end{align*}
where 
\[ \und U(\mu) = (1 - \delta)\sum_{\xi = 0}^\infty \delta^\xi \bb{E}_{M^\xi\mu}[u(a^*(M^\xi\mu, \textbf{0}), \theta)] \text{  with  } a^*(\tilde \mu, \textbf{0}) \in \argmax_{a \in A} \bb{E}_{\tilde \mu} [u(a, \theta)]\]
is Receiver's expected no-information payoff, i.e. the value of their outside option. 

The first two constraints (incentive compatibility and promise keeping) are standard in dynamic contracting, noting the Markovian evolution of the prior given by the belief (and the choice of $\rho$). The third constraint, Bayes plausibility, is unique to the persuasion problem. The final constraint, boundedness, is a technical assumption that ensures solutions for $V$ satisfy a baseline transversality condition that the agent is not ``strung-along forever,'' i.e. promised infinite payoff at infinity but never at any finite time. In practice, $C$ can be quite large (i.e. if $|u|$ is bounded by $1$, $C$ can be $1000^{1000}$), and the specific choice of $C$ will not affect the characterization or main results. Throughout, suppose $C$ is at least greater than the total possible surplus in the problem. 
When there is no loss of ambiguity, I will omit the dependence of $V$ on $\delta$ (as often throughout the paper I implicitly fix a discount rate). 

Proposition \ref{p: FE SP equivalence} formalizes the way in which solving for the functional equation above implies finding an optimal $\sigma$. 
Throughout, represent $\sigma^S(h^\xi)$ as a tuple $\{\rho(h^\xi), t(h^\xi)\}$ and Receiver action $\sigma^R(h^\xi)$ as a random action $\{\alpha(h^\xi)\}$, and let $\mcal U((h^\xi, \mu, a); \sigma)$ be Receiver's continuation utility at history $h^\xi$ after forming posterior belief $\mu$ and taking action $a$ (both assumed to be on the path of play). When there is any ambiguity about which contract $\{\rho(\cdot), t(\cdot), \alpha(\cdot)\}$ represents, I will add an additional argument to specify explicitly, i.e. $\{\rho(\cdot; \sigma), t(\cdot; \sigma), \alpha(\cdot; \sigma)\}$ though I suppress the extra notation in the absence of such ambiguity. 

\begin{proposition}
    \label{p: FE SP equivalence}
(\ref{equation_1}) gives the value of Sender's problem: $V(0, \mu_0, \delta) = \mcal V^*(\mu_0, \delta)$. 
    Moreover, there exists an optimal $\sigma$ where for any $h^\xi \in \supp(\bb{P}^\sigma)$, 
    \[ \{\rho(h^\xi), t(h^\xi), \{\mcal U(\{h^\xi, \mu, a\})\}_{\mu \in \supp(\rho(h^\xi)), a \in \supp(\alpha(h^\xi)(\mu))}, \alpha(h^\xi)\} \]
    solves (\ref{equation_1}) at $\mcal U(h^\xi; \sigma)$. 
\end{proposition}

Proposition \ref{p: FE SP equivalence} follows from basic principles in dynamic programming and is proven in \blue{\ref{p: FE SP equivalence proof}}
Importantly, it implies that it is sufficient to characterize (1) the evolution of the argument maximizers to the functional equation at different prior beliefs and utility promises and (2) the evolution of the path of utility promises at the optimum in order to obtain a full characterization of the optimal contract $\sigma^*$. Towards this end, let $\mcal F(u, \mu)$ be the collection of all tuples which solve the value function program (for an implicit fixed discount rate) at utility promise $u$ and belief $\mu$. 
I conclude the description of the model by stating some basic facts about the value function. 

\begin{proposition}
\label{p: FE properties}
    There exists a unique $V$ solving (\ref{equation_1}), and moreover $V$ is continuous in $u$ over $[-C, C]$ and $\mu_0$ on $\Delta(\Theta)$. Finally, $V$ is concave nonincreasing in $u$, and its right derivative satisfies $V_+'(u, \mu) \geq -k$ for any $u, \mu$. 
\end{proposition}

The proof follows standard arguments in dynamic programming (see Chapter 4 of \cite{StokeyLucasPrescott1989}). For completeness, I outline the proof in \blue{\ref{p: FE properties proof}}

\section{Time Structure of Payments}

The ability to provide information augments dynamic contracts in two ways. First is a \emph{static persuasion} effect: it changes the value of any action taken today because Receiver's expected state-dependent utility at the realized posterior belief changes. Second is a \emph{dynamic incentivization} effect: because Receiver values more accurate information, Sender can leverage their ability to provide future information as a way to incentivize effort today. 
This tradeoff leads to several novel questions inherent to the model. How does this incentivization effect compare to transfers in effectiveness? How does Sender balance static persuasion with dynamic incentivization? Does Sender ever want to link Receiver's dynamic incentives---i.e. promise more information tomorrow (even when doing so might be beneficial to Receiver at the cost of Sender's continuation utility) as a reward for taking less favorable actions today?

A sharp and simple class of mechanisms that cleanly resolves the above questions are \emph{backloaded} mechanisms---those where payments occur only after there is no room for information provision as a dynamic tool. Intuitively, these mechanisms say that whenever Sender wants to link Receiver's dynamic incentives, doing so via information---giving up static persuasion power in favor of dynamic incentivization---is (weakly) more efficient than just paying Receiver directly. Formally, I define these mechanisms as follows, where an information structure $\rho$ \emph{reveals the state} if it only supports degenerate beliefs:

\begin{definition}
   \emph{Transfers are a last resort} at $\sigma$ if, for any $h^\xi \in \supp(\bb{P}^\sigma)$, $t(h^\xi)(\mu, a) > 0$ implies $\rho(h^\zeta)$ reveals the state at every on-path successor history $h^\zeta \succsim \{h^\xi, \mu, a\}$. 
\end{definition}

\noindent Perhaps surprisingly, there is always an optimal mechanism where transfers are a last resort.  

\begin{theorem}
\label{t: transfers as last resort}
There is an optimal contract $\sigma^*$ where transfers are a last resort. 
\end{theorem}

Theorem \ref{t: transfers as last resort} is the main takeaway of this paper. It gives a characterization of when payments should be used to dynamically motivate Receiver in the presence of informational incentives. If Sender ever wants to motivate an agent via transfers, they should instead turn to dynamic information first. Along the optimal contract where transfers are a last resort, Sender should start using transfers to motivate the agent only after they have completely ``drawn down'' on their stock of information (by promising full information at all continuation histories).

The intuition behind the result is as follows. Suppose Sender at any point wishes to motivate the agent to take an action (fixing an induced belief today). They can do this in two ways: by paying the agent, or by promising them more information in the future. Since more information makes Receiver's decision more efficient, Sender can always give a little more information to increase Receiver's utility (and possibly Sender's at some beliefs). From here, at any of the newly induced posterior beliefs where Receiver's new action adversely affects Sender's utility, they can pay Receiver to take an action equal to their original action. The amount they have to pay is exactly equal to the change in Receiver's payoff along this new posterior relative to the new action, and thus comes at a efficiency cost of at most rate $k$. 
Such a future joint pay-and-persuade scheme cannot be worse off than simply paying the agent today (and will do strictly better so long as more information benefits Sender too), and hence so long as such a scheme is feasible at some future continuation history, can be chosen in lieu of payments today along the optimal path. 
This observation is the substance of the pullback and squeezing lemmas (Lemmas \ref{l: pullback lemma} and \ref{l: squeezing lemma}). The remainder of the proof handles the technicalities that arise given the generality of the model and the infinite set of histories. 

It is worth explicitly flagging that the above informal argument is distinct from two alternative intuitions one might have about the result. First is the intuition that transfers must be backloaded because information provision is ``always free'' (since there are no costs to information acquisition). In fact, it can be that in the absence of transfers, the cost of giving the agent more information can be arbitrarily high because it allows Receiver to take actions which can be very damaging to Sender but which only marginally benefit Receiver.
With transfers, however, the pullback lemma shows that Sender can bound the cost of transferring utility to Receiver via information and transfers by at most $k$, which is the marginal cost of simply paying an agent to fulfill a utility promise.
Second is the intuition that transfers can be mechanically backloaded because payments are perfectly intertemporally substitutable. While it is true payments can be mechanically pushed backwards (at the cost of increasing them by $\frac{1}{\delta}$) while maintaining optimality, this mechanical push-back does not change the information structure. Thus, there is no guarantee after pushing back payments that (1) for all optimal contracts the state is fully revealed or that (2) pushing transfers back would encourage information revelation. Thus, there is no way to guarantee by simply backloading transfers that they would ever be used as a last resort in some optimal contract. In contrast, my construction (in Lemma \ref{l: squeezing lemma}) (1) pushes transfers into specific future histories where the state is not fully revealed, and (2) simultaneously reveals more information to increase the agents' indirect consumption utility. Combined, this alleviates the need for transfers today, while also guaranteeing that future histories see more informative experiments. Finally, transfers are guaranteed to be a last resort by appealing to an appropriate compactness argument, after having pushed utility far enough back that all future experiments fully reveal the state (Lemma \ref{l: squeezing lemma}). 

Moreover, I remark here that Theorem \ref{t: transfers as last resort} is conceptually distinct from familiar backloading results in the dynamic contracting literature (\cite{ray2003time}, \cite{guo2020dynamic}). First, it is not a pure incentive backloading result---informational incentives can still arrive over time. Second, it is a sequencing argument between payments and information, not a characterization of optimal asymptotic contracts. Third, unlike \cite{ray2003time}, who requires discount factors are not too high\footnote{This prevents Sender from obtaining first best; see Assumption (A.5) of \cite{ray2003time}.}, my characterization holds for all $\delta$ (and in fact gives nontrivial payoff predictions as $\delta \to 1$: see Proposition \ref{p: persuasion upper bound}). Finally, even with quasilinear transfers, the asymptotic contract neither needs to (1) immiserate Receiver, or (2) settle into the one which maximizes Receiver payoffs. 

There are two technical challenges that make the proof and structure of Theorem \ref{t: transfers as last resort} distinct from familiar arguments in the dynamic contracting literature. 
First, when Sender can persuade, their choice of information structure endogenously changes Receiver's expected utility of each action and their outside option, leading to additional subtleties which are absent in standard principal-agent interactions. These interactions are first-order: Lemma \ref{l: squeezing lemma} critically shows that the cost of mollifying Receiver's outside option by providing information is bounded from above by the cost of directly paying Receiver. 
Second, my limited liability constraint holds \textit{pathwise}, and thus can be more complicated. This is because an optimal experiment may induce beliefs both where the limited liability constraint binds (i.e. incentive are aligned, so Sender would like to be paid by Receiver) and where Sender is paying Receiver to make them exactly indifferent between their favorite action and a different one. This asymmetry again shows up in the argument behind Lemma \ref{l: squeezing lemma}: Sender can exactly increase the probability of generating the ``more profitable belief'' in a way that makes Receiver better off as a way to compensate Receiver \textit{in lieu of} giving them money. 

Several assumptions are not necessary for Theorem \ref{t: transfers as last resort}. For example, it need not be that the payoffs for players is the same in every period---payoffs can evolve in any Markovian fashion as well, so long Receiver's payoff function is known to themselves at the start of each period.\footnote{For example, in the motivating rideshare example, the driver's value for rides may vary with each draw.} Moreover, the state can be (partially, or noisily) revealed at the end of each period, so long as the revelation or monitoring structure is independent of Sender's action. The proof will make clear why these assumptions are not necessary, but also why it would be notationally cumbersome to directly accommodate them. 

Finally, note Theorem \ref{t: transfers as last resort} does not guarantee every optimal contract strictly backloads transfers; only that one such contract exists. In Sections 4.3 and 4.4 (see Propositions \ref{p: weak ray regularity} and \ref{p: always last resort}), I expand on this indeterminancy and give conditions under which transfers are strictly backloaded---there are always periods where dynamic information design is strictly preferred to transfers as a way of motivating Receiver.  

I conclude this section with the proof of Theorem \ref{t: transfers as last resort}.
\begin{proof}
 The first step is to characterize the slope of $V(\cdot)$ when transfers are used. 

\begin{lemma}
    \label{l: slope of value at optimum}
 For any $(u, \mu_0)$ and  $\{\rho, t, u', \alpha\} \in \mcal F(u, \mu_0)$:
    \begin{enumerate}
        \item If $V_+'(u, \mu_0) = -k$, $V_+'(u'(\mu, a), \mu) = -k$ for all $\mu \in \supp(\rho)$, $a \in \supp(\alpha(\mu))$. 
        \item If $t(\mu, a) > 0$, then $V_+'(u'(\mu, a), \mu) = -k$ for all $\mu \in \supp(\rho), a \in \supp(\alpha(\mu))$.  
    \end{enumerate}
\end{lemma}

The proof can be found in the \blue{\ref{l: slope of value at optimum proof}}
Lemma \ref{l: slope of value at optimum} first shows that once transfers are used, then Sender and Receiver must be at the steepest part of the Pareto frontier forever more in the relationship. Moreover, it shows that if payments ever occur, then this must be true. These observations imply the steepest part of the Pareto frontier is absorbing, and being at the steepest part of the slope is a necessary condition for using transfers. 

The second step: once we are at the steepest part of the Pareto frontier, it is without loss of optimality to provide incentives by giving Receiver full information with some probability.
To state this step we need a bit more notation. 
At any time $\xi$, let $\bb{Q}_\xi^\sigma \in \Delta(\Theta \times A)$ be the \textit{unconditional} probability distribution over outcomes  (averaging over histories) induced by contract $\sigma$ at time $\xi$, with $\bb{Q}_\xi^\sigma(\theta) \in \Delta(A)$ the distribution over actions fixing some state. Moreover, define 
\[ u^{RFI}(\mu) = (1 - \delta)\sum_{\xi = 0}^\infty \delta^\xi  \bb{E}_{M^\xi \mu} \left[\max_{a \in A} u(a, \theta)\right]\]
to be Receiver's expected full-information payoff, starting at belief $\mu$. 
Finally, say $\sigma$ is $u$-constrained optimal if it is optimal among all obedient $\sigma$ which guarantee Receiver a payoff of at least $u$. Let $mu(h^\xi, \sigma)$ be the posterior belief over states given history $h^\xi$ and strategy $\sigma$. Note all continuation contracts $\sigma(\cdot | h^\xi)$ are $\mcal U(h^\xi | \sigma)$-constrained optimal. 
We can now state the following lemma, proven in the \blue{\ref{l: pullback lemma proof}} 

\begin{lemma}[Pullback Lemma]
\label{l: pullback lemma}
    Fix $(u, \mu_0) = (\mcal U(h^\xi | \sigma), \mu(h^\xi, \sigma))$ such that $V_+'(u, \mu_0) = -k$, and let $\sigma$ be $u$-constrained optimal. Then there exists obedient $\sigma^*$ such that 
    \begin{enumerate}
        \item $\rho(h^\zeta; \sigma^*)$ reveals the state for all $h^\zeta \succ h^\xi$, $h^\zeta \in \supp(\bb{P}^{\sigma^*})$. 
        \item At every time $\xi$, $\zeta \geq \xi$, $\bb{Q}_\zeta^\sigma(\cdot | h^\xi) = \bb{Q}_\zeta^{\sigma^*}(\cdot | h^\xi)$. 
        \item $\mcal V(h^\xi | \sigma) - \mcal V(h^\xi | \sigma^*) \leq k(u^{RFI}(\mu_0) - \mcal U(\sigma))$. 
        \item For every $\tilde u \in [u, u^{RFI}(\mu_0)]$, there exists $\alpha \in [0, 1]$ such that $\alpha \sigma^* + (1 - \alpha)\sigma$ is $\tilde u$-constrained optimal. 
    \end{enumerate}
\end{lemma}

The pullback lemma combined with Lemma \ref{l: slope of value at optimum} implies that at any history where transfers are used, it is possible to further increase utility by simply revealing the state with positive probability. 
This comes at a cost of (at most) $k$, and hence is an efficient way to transfer utility once $V_+'(u, \mu_0) = -k$. This implies that full information is on the Pareto frontier at some point, since it is always efficient to provide utility by giving a little bit more information (i.e. an increase in $\tilde u$ corresponds to an increase in $\alpha$). 

The third step. From here, fix an optimal $\sigma^*$ and define the set of histories where transfers are not a last resort, $\mcal H^\xi(\sigma^*)$, to be:
\begin{align*}  
\mcal H^\xi(\sigma^*) = \{h^\xi \in \supp(\bb{P}^{\sigma^*}) \cap H^\xi:  \text{ } & t(h^\xi)(\mu, a) > 0, \mu \in \supp(\rho(h^\xi; \sigma^*)), a \in \supp(\alpha(h^\xi; \sigma^*)(\mu))
\\ & \text{ but } \exists h^\zeta \succ (h^\xi, \mu), h^\zeta \in \supp(\bb{P}^{\sigma^*}) \text{  s.t.  } \rho(h^\zeta; \sigma^*) \neq \rho^{FI}\} 
\end{align*}
where $\rho^{FI}$ is the experiment that always reveals the state. 

\begin{lemma}[Squeezing Lemma]
\label{l: squeezing lemma}
    Fix optimal $\sigma^*$ and $h^\zeta \in \mcal H^\zeta(\sigma^*)$. Then there exists optimal $\bar \sigma$ satisfying $\bb{P}_\phi^{\sigma^*} = \bb{P}_\phi^{\bar \sigma}$ for all $\phi \leq \zeta$ such that $h^\zeta \not\in \mcal H^\zeta(\bar \sigma)$. 
\end{lemma} 

The proof can be found in \blue{\ref{l: squeezing lemma proof}}
The pullback and squeezing lemmas give an iterative procedure we can use to obtain an optimal contract where transfers are a last resort.  
Start with some optimum $\sigma_0$. Define the function
\[ \varphi(\sigma) = \inf\{t : \mcal H^\xi(\sigma) \neq \varnothing\}. \]
If $\varphi(\sigma_0) < \infty$, start from $\varphi(\sigma_0)$ and apply the squeezing lemma to each history $h^{\varphi(\sigma_0)} \in \mcal H^{\varphi(\sigma_0)}(\sigma_0)$. This induces optimal $\sigma_1$ where $\varphi(\sigma_1) > \varphi(\sigma_0)$. If now $\varphi(\sigma_1) < \infty$, again apply the squeezing lemma to all $h^{\varphi(\sigma_1)} \in \mcal H^{\varphi(\sigma_1)}(\sigma_1)$. Continuing, one of two things must happen.
\begin{enumerate}
    \item For some finite $N$, $\varphi(\sigma_N) = \infty$. Then $\sigma_N$ is an optimal contract where transfers are a last resort, and we are done. 
    \item There is a strictly increasing sequence $\{\varphi(\sigma_k)\}_{k = 1}^\infty$ where $\{\sigma_k\}_{k = 1}^\infty$ are optimal. 
\end{enumerate}

In the latter case, a compactness argument gives a convergent subsequence $\{\sigma_{k_n}\} \subset \{\sigma_k\}$ in the strong topology. Let $\sigma_\infty$ be this subsequential limit. Since optimal contracts are closed (because $\bb{P}^\sigma$ is continuous in $\sigma$ by \blue{\ref{Appendix B}}), $\sigma_\infty$ is also optimal. But $\varphi(\sigma_\infty) = \infty$ must be true, and so $\sigma_\infty$ is an optimal contract where transfers are a last resort. 
\end{proof}

\section{Optimal Contracts}
This section gives a partial characterization of optimal contracts by deriving properties of optimality which hold uniformly across all optimum. Notably, this differs from the sharp characterization of optimal contracting given in Theorem \ref{t: transfers as last resort}, which states there \textit{exists} an optimal contract with the property that transfers are a last resort. 

\subsection{Feasible Optimality} 

I start by giving a necessary condition for an action to be induced on the path of play for any optimal contract. The following is the key condition, which tracks the benefit to Sender of inducing that action against the cost they must incur if they were to use transfers to incentivize Receiver to take that action. 

\begin{definition}
    Define the \textit{feasibly optimal} set $\mathscr{F}$ to be the set 
    \[ \left\{a : \not\exists a' \neq a, \min_{\theta \in \Theta} \left\{ v(a', \theta) - v(a, \theta)\right\} > k \max_{\theta} \left\{u(a, \theta) - u(a', \theta)\right\} \right\}. \]
\end{definition}

Rearranging and taking expectations, an action $a$ is feasibly optimal if, for all $\mu$, the marginal rate of substitution between Sender and Receiver preferences is at least $-k$; that is, for all beliefs $\mu$, 
\[ \bb{E}_\mu[v(a', \theta) - v(a, \theta)] > -k{\bb{E}_\mu[u(a', \theta) - u(a, \theta)]}\] 
Feasibly optimal actions are named exactly because they are the only actions which can appear on the path of play in \textit{any} optimal contract. 

\begin{proposition}
    \label{p: feasibly optimal actions}
    Suppose $h^\xi \in \supp(\bb{P}^\sigma)$ for optimal $\sigma$. Then for all $a \in \alpha(h^\xi)(\mu)$, $\mu \in \supp(\rho(h^\xi))$, $a \in \mathscr{F}$. 
\end{proposition}

The formal proof is relegated to the \blue{\ref{p: feasibly optimal actions proof}} 
The intuition is as follows. Suppose $a$ is not in the feasibly optimal set, and let $a'$ be some action which ``beats'' it (i.e. keeps it from being in $\mathscr{F}$). Then whenever Sender would induce $a$ as part of an optimal contract, they could do strictly better by paying the agent to take action $a'$ instead, regardless of the posterior belief induced when $a$ is taken. 

Proposition \ref{p: feasibly optimal actions} is easy to check in the case with state-independent actions.   

\begin{corollary}
    \label{cor: state indep actions}
    Let $a_1 = \argmax_{a \in A} v(a)$ be Sender's favorite action and suppose $v(a_1) - v(a) > k\max_\theta\{u(a, \theta) - u(a_1, \theta)\}$ for all $a \in A \setminus \{a_1\}$. Then at every $h^\xi$, $\alpha(h^\xi)(\mu) = \delta_{a_1}$. 
\end{corollary}

Corollary \ref{cor: state indep actions} thus gives a unique prediction when Sender has state-independent preferences and $k$ is sufficiently low. 
While straightforward given Proposition \ref{p: feasibly optimal actions}, it will simplify the characterization of the optimal contract in the leading example (Theorem \ref{t: loyalty contracts}). 
In general, the feasibly optimal set will also play a role in characterizing the set of beliefs which can arise on the path of play at any optimum as well, as I next characterize.

\subsection{Effectiveness Ratios}
Theorem \ref{t: transfers as last resort} should be interpreted as a partial characterization of optimal contracting: there \textit{exists} an optimum where transfers are a last resort.
However, this need not be the only optimal contract. This is because dynamic incentives are perfectly substitutable along the Pareto frontier whenever it is linear (as providing $1$ unit of incentives today is as expensive as providing $\frac{1}{\delta}$ units tomorrow), and thus information and transfers are perfectly substitutable whenever $V_+'(u, \mu) = -k$ and $u \leq u^{RFI}(\mu)$. 
Thus, it could be that beliefs which are nondegenerate are induced after transfers in \textit{some} optimum. This subsection gives an answer to the following uniform strengthening of Theorem \ref{t: transfers as last resort}: which beliefs are inconsistent with optimality at any continuation profile following a history where transfers have been used? To state the result, it will be helpful to specialize to the i.i.d. case, and make the following mild (generic) assumption on payoffs to ensure Receiver always values receiving information. The following assumption rules out the case where Receiver is indifferent between all information structures, in which case information is indeterminate in motivating Receiver.  

\begin{definition}
\label{d: effectiveness ratio}
\emph{Full-information is strictly optimal for Receiver} if  
\[ \bb{E}_{\rho^{FI}}\left[ \bb{E}_\mu[u(a^*(\mu, \mathbf{0}), \theta)] \right] > \bb{E}_{\rho}\left[ \bb{E}_\mu[u(a^*(\mu, \mathbf{0}), \theta)] \right] \]
for any $\rho \neq \rho^{FI}$. 
\end{definition}

Definition \ref{d: effectiveness ratio} is satisfied if, for example, Receiver's full-information best-response correspondence $a^*(\theta)$ is single-valued and injective in $\theta$. 
This definition guarantees the following \textit{effectiveness ratio}, which will play a crucial role in the characterization, is well-defined. 

\begin{definition}
    Suppose full information is strictly optimal for Receiver. For any belief $\mu$, define the \emph{effectiveness ratio} of $\mu$ to be  
\[ E(\mu) =  \min_{a' \in \mathscr{F}}\left\{\frac{\bb{E}_\mu[v(a^*(\delta_\theta, \mathbf{0}), \theta) - v(a', \theta)]}{\bb{E}_\mu[u(a^*(\delta_\theta, \mathbf{0}), \theta) - u(a^*(\mu, \mathbf{0}), \theta)]}\right\}. \]
\end{definition}

Let $\mcal E(k) = \{\mu \in \Delta(\Theta)^o : E(\mu) > -k\}$ be the set of all beliefs whose effectiveness ratio is strictly greater than $-k$. 
Here, $a^*(\mu, \mathbf{0}) \in \argmax_{a \in A} \bb{E}_\mu[u(a, \theta)]$, abusing notation to refer to the correspondence as a function when there is no loss of ambiguity.

\begin{proposition}
    \label{p: strict optimality}
    Suppose 
    full-information is strictly optimal for Receiver. Then for any optimal $\sigma$ and $(h^\xi, \mu, a) \in \bb{P}^\sigma$, if $t(h^\xi; \sigma)(\mu, a) > 0$, then $\supp(\rho(h^\zeta; \sigma)) \subset \mcal E(k)^c$ for all $h^\zeta \succ (h^\xi, \mu, a)$. 
\end{proposition}

The formal proof is in the \blue{\ref{p: strict optimality proof}}
The intuition for the result is as follows.
Fix some belief $\tilde \mu \in \mcal E(k)$. By definition of the effectiveness ratio, Sender's loss from revealing the state instead of inducing $\tilde \mu$ and letting Receiver take their favorite action is less than $k$ times Receiver's gain. 
Should Sender then need to dynamically motivate Receiver, they would rather split $\tilde \mu$ into its constituent parts than pay Receiver, as the latter carries a cost of $-k$. Consequently, the presence of a belief $\tilde \mu$ is inconsistent with the fact that transfers could possibly be used. Geometrically, Proposition \ref{p: strict optimality} implies the Pareto frontier at any utility promise where $\tilde \mu$ may be supported by an optimal experiment has slope strictly greater than $-k$, and hence by Lemma \ref{l: slope of value at optimum}, transfers cannot be used. 

Proposition \ref{p: strict optimality} shows the effectiveness ratio of a belief is sufficient to rule out the belief being induced at any optimum where transfers are used. The next proposition gives a geometric interpretation of these sets of beliefs as a function of the effectiveness of transfers. 

\begin{proposition}
\label{p: Ek convex}
    $\mcal E(k)$ is convex and $\mcal E(k') \subset \mcal E(k)$ whenever $k' < k$. 
\end{proposition} 
The proof follows from some straightforward computations (these can be found in \blue{\ref{p: Ek convex proof}}) I interpret Proposition \ref{p: Ek convex} as a comparative statics result: as transfers become relatively more expensive, the set of beliefs that are ruled out strictly expands since $\supp(\rho(h^\zeta; \sigma)) \subset \mcal E(k)^c$, so that only those beliefs which are ``more extremal'' remain consistent with optimality after payments. As $k \to \infty$, the degenerate beliefs (i.e. the extreme points of the probability simplex) become the only beliefs consistent with payments. This is because Sender can always, in lieu of payments, reveal the state and have Receiver take their favorite action, instead of using transfers to change Receiver actions.

\subsection{Incentivizability}
One way to interpret Theorem \ref{t: transfers as last resort} is that there exists some history-dependent stopping time $T_\sigma^*$ such that at each history, before $T_\sigma^*$, only (dynamic) informational incentives are used to motivate Receiver, while after $T_\sigma^*$, only transfers are used as a way to motivate Receiver. 
Theorem \ref{t: transfers as last resort} leaves open the question of whether Sender finds it strictly profitable to sequence information and transfers. In particular, one might worry that the time structure of optimal information and transfers holds only because \textit{all} combinations of information and transfers are optimal, and hence Theorem \ref{t: transfers as last resort} merely isolates one such contract. However, this is not the case---this subsection identifies conditions that guarantee dynamic informational incentives are used strictly before transfers in any optimum. Moreover, these conditions are natural: they are satisfied whenever repeated static persuasion is not optimal in the class of all stationary contracts.\footnote{
In \blue{\ref{Appendix C}}, I show that in sufficiently continuous problems, the regularity conditions identified here hold for a broad class of economic primitives.} 
Proposition \ref{p: weak ray regularity} thus highlights the importance of (strictly) frontloading informational incentives before paying Receiver to motivate them. 

I suppose throughout this subsection the state is i.i.d. and normalize payoffs so that $\und U(\mu_0) = 0$.\footnote{This assumption is made for expositional convenience and is not necessary for Proposition \ref{p: weak ray regularity}.} 
I adopt the following definition for when, across all optimum, informational incentives must take strict precedence over monetary ones. 

\begin{definition}
    \label{d: transfers strictly backloaded}
    Transfers are \emph{strictly backloaded} at prior $\mu_0$ if every $\{\rho, t, u', \alpha\} \in \mcal F(0, \mu_0)$ has $u'(\mu, a) > 0$ for some $\mu \in \supp(\rho)$ and $a \in \supp(\alpha(\mu))$. 
\end{definition}

Why does Definition \ref{d: transfers strictly backloaded} capture this notion? Suppose instead that there was an optimal contract which did not use informational incentives prior to transfers. There are two cases. First is if the utility promise never jumps up, in which case $u'(\mu, a) = 0$ always and transfers are not strictly backloaded according to Definition \ref{d: transfers strictly backloaded}. The second case is if the utility promise does jump at some point $u'(\mu, a) > 0$ but is fulfilled using only transfers at some future date. Since transfers are perfectly substitutable across time, it is without loss of optimality to frontload those transfers to the first period---to pay $\frac{\delta}{1 - \delta} u'(\mu, a)$ today instead of $u'(\mu, a)$ in future payments. But then this implies the existence of an optimal contract that pays today instead of in the future, guaranteeing Definition \ref{d: transfers strictly backloaded} fails. Thus Definition \ref{d: transfers strictly backloaded} exactly captures when the use of some informational incentives must be strictly preferred by Sender, as opposed to using payments directly. 

What are the conditions under which transfers are strictly backloaded at every optimum? Two conditions are clearly necessary.
First, it must be that the Pareto frontier is not flat, i.e. $V_+'(0, \mu_0) \neq -k$. If this were the case, then Lemma \ref{l: pullback lemma} would imply transfers and information were exactly perfect substitutes, and hence any combination of (pulled back) information and transfers would be part of an optimal way to move along the Pareto frontier. 
Second, there must be some role for dynamic incentives, so that a stationary optimum where Receiver always takes $a^*(\mu, \mathbf{0})$ is not always optimal. If this were the case, then there would be no role for dynamic incentives, either through transfers \textit{or} information. 
The next result shows these two clearly necessary conditions are in, fact, exactly sufficient: they characterize when transfers are strictly backloaded. 

\begin{definition}
    \label{d: nontriviality}
    Primitives $\{u, v, k, \delta\}$ are \emph{nontrivial} at prior $\mu_0$ if $V_+'(0, \mu_0) > -k$. They are \emph{incentivizable} at $\mu_0$ if for each $\{\rho, t, u', \alpha\} \in \mcal F(0, \mu_0)$, there exists $\mu \in \supp(\rho)$ and $a \in \supp(\alpha(\mu))$ such that $a \neq a^*(\mu, \mathbf{0})$. 
\end{definition}

\begin{proposition}
    \label{p: weak ray regularity}
    Transfers are strictly backloaded at $\mu_0$ if and only if primitives  $\{u, v, k, \delta\}$ are nontrivial and incentivizable at $\mu_0$. 
\end{proposition}

Proposition \ref{p: weak ray regularity} strengthens Theorem \ref{t: transfers as last resort} in the following way: it provides a history-dependent stopping time $T_\sigma^*$, before which only informational incentives are used to motivate Receiver (and indeed they are used), and after which transfers can potentially be used. 
The proof can be found in the \blue{\ref{p: weak ray regularity proof}}
Nontriviality is the easier of the two conditions to satisfy, and in particular will hold so long as $k$ is sufficiently large (i.e. transfers are expensive enough that information is ``cheaper'' at a first pass). Incentivizability is the less obvious of the two conditions to be satisfied, since it requires control over every solution to the dynamic contracting problem, which is itself an endogenous equilibrium object. I conclude this section by giving an easier to check sufficient condition for incentivizability. 

For the statement of the next result, let $\mcal V^*(\mu_0, 0)$ be the value to the static problem, and $\text{cav}(V_0)(\mu_0)$ to be the solution to the static problem \textit{without transfers.} Note that $\mcal V^*(\mu_0, 0) > \text{cav}(V_0)(\mu_0)$ if and only if transfers are useful to Sender in the static one-shot problem. Proposition \ref{p: wrr if static k-cavification} shows that so long as transfers are useful in the static one-shot problem, then the problem is incentivizable.  

\begin{proposition}
    \label{p: wrr if static k-cavification} 
    Primitives $\{u, v, k, \delta\}$ are incentivizable if $\mcal V^*(\mu_0, 0) > \text{cav}(V_0)(\mu_0)$. 
\end{proposition}

Note Proposition \ref{p: wrr if static k-cavification} gives a condition for incentivizability for \textit{any} $\delta > 0$, as a function of behavior at $\delta = 0$. A formal proof is in \blue{\ref{p: wrr if static k-cavification proof}} To see why the result is true, note that if the problem is not incentivizable, then there is an optimum where $a = a^*(\mu, \mathbf{0})$ always on the path of play. But this solution, which is feasible in the information design problem without transfers, must be optimal for the dynamic contract where $\delta = 0$ as well, since utility promises and intertemporal transfers are never used. Hence $\mcal V^*(\mu_0, 0) = \text{cav}(V_0)(\mu_0)$. 

One objection to Proposition \ref{p: wrr if static k-cavification} is that verifying whether transfers are used in the static problem might itself be a computationally difficult exercise, and hence Proposition \ref{p: wrr if static k-cavification} simply punts one relatively inscrutable condition (incentivizability) into a different one. Luckily, this is not the case; in Section 5.1, I develop a simple algorithm that verifies whether transfers are used in the joint static problem when $\delta = 0$ by looking at ``maximally indifferent'' beliefs. 

A combination of the algorithm developed next and Proposition \ref{p: wrr if static k-cavification} makes incentivizability much easier to check. For example, the class of problems studied in Section 6 are incentivizable whenever there exists some $c_i$ such that $c_i > k$. Alternatively, the repeated version of the example given in \blue{\ref{Appendix D}} satisfies incentivizability, so strictly backloads transfers. 

\subsection{Expensive Transfers}

The previous subsection gave conditions under which periods of pure information design are strictly preferred to using transfers in some periods of every optimal contract; that is, there are periods where dynamic information is strictly preferred to transfers as a motivating tool. This shows that optimal contracts consists of two nontrivial phases---one with only information, and one including transfers. However, it does not show every optimal contract must feature transfers as a last resort. 
In this subsection, I show that when transfers become sufficiently expensive (in the sense that $k$ is sufficiently large), then \emph{every} optimal contract must feature transfers as a last resort. Formally, 

\begin{definition}
    Transfers are \emph{always a last resort} if for every optimal contract $\sigma^*$, transfers are a last resort at $\sigma^*$. 
\end{definition}

\begin{proposition}
    \label{p: always last resort}
    Fix any $\{u, v, \delta\}$. There exists some $\bar k \in \bb{R}_+$ such that for all $k > \bar k$, transfers are always a last resort at primitives $\{u, v, \delta, k\}$. 
\end{proposition}

A formal proof is given in \blue{\ref{p: always last resort proof}} 
The intuition behind the proof is best read in contrast to the proof of Theorem \ref{t: transfers as last resort}. Recall the pullback lemma (Lemma \ref{l: pullback lemma}) states that it is always possible to jointly provide information and transfers in the future at a rate of at most $-k$. However, in the absence of transfers, there is a rate of providing future incentives using only information that depends only on $(u, v, \delta)$. So long as $k$ is larger than some bound on this rate, then the pullback lemma never needs to operate, since providing information as an incentive is always cheaper than jointly paying and persuading. Thus, every optimal contract will feature transfers as a last resort, using an argument akin to the squeezing lemma. 

Three remarks are in order. First, the result is more subtle than simply stating that as $k \to \infty$, transfers are never used. For example, in the setting of Corollary \ref{cor: state indep actions} with high stakes (i.e. $v(a_1) - v(a)$ is large), transfers will eventually be used on the path of play to ensure a constant action is induced even as they get more expensive. Hence transfers will be a last resort at every optimum as $k$ gets large, but will still be used on-path. 
Second, these conditions are complementary to, but do not subsume, the analysis in the previous two sections. In particular, the previous analysis held for any choice of primitives $\{u, v, \delta, k\}$, while this new analysis holds in the limit as $k \to \infty$. Finally, the result contrasts naturally with Theorem \ref{t: transfers as last resort}; while I require here that $k$ is large enough that the pullback argument is essentially trivial, Theorem \ref{t: transfers as last resort} holds \emph{regardless of how cheap transfer are}. This perhaps highlights the surprising nature of Theorem \ref{t: transfers as last resort}, as transfers are used as a secondary incentive instrument to information even as they become arbitrarily cheap.

\section{Valuing Persuasion}
\subsection{The Myopic Case}
Consider the case where $\delta = 0$ so Sender and Receiver are fully myopic. In this case, it is without loss of generality to focus on only the first period, and so Sender's contract is simply a transfer function $t: \Delta(\Theta) \times A \to \bb{R}_+$ and an experiment $\rho \in \Delta(\Delta(\Theta))$ which is $\mu_0$-Bayes plausible. For any transfer rule $t$, define Receiver's best response (breaking ties in favor of Sender) as 
\[ a^*(\mu, t) = \argmax_{a \in a^\dagger(\mu, t)} \left\{ \bb{E}_\mu[v(a, \theta) - k t(\mu, a) \right\} \text{  where  } a^\dagger(\mu, t) = \argmax_{a \in A} \left\{ \bb{E}_\mu[u(a, \theta) + t(\mu, a)] \right\} \]
where I appeal to Lemma \ref{l: transfers not depend on state} to simplify the space of transfers. As before, let $a^*(\mu, \mathbf{0})$ to be Receiver's optimal action in the absence of transfers.

Sender's problem is now 
\[ \max_{\rho \in \Delta_{\mu_0}(\Delta(\Theta)), t} \left\{\bb{E}_\rho\left[ \bb{E}_\mu[v(a^*(\mu, t), \theta) - k t(\mu, a^*(\mu, t)))] \right] \right\}. \]
Maximizing first over transfers and then information policies, and applying the concavification theorem of \cite{KamenicaGentzkow11} implies Sender's value from persuasion can be written as 
\[ V^*(\mu_0) = \text{cav}|_{\mu_0}\left( \max_t \bb{E}_\mu[v(a^*(\mu, t), \theta) - k t(\mu, a^*(\mu, t)))]  \right)\]
where transfers are maximized belief-by-belief and the concavification is evaluated at $\mu_0$. This however can still be a potentially complicated object, because the optimal transfer rule must be found belief-by-belief, and the resulting indirect utility function will need to be concavified (which is itself potentially a difficult procedure). 
It turns out however, that the linearity of transfers gives a simple characterization of $V^*(\mu_0)$ as the linear interpolation of finitely many points. To state it, I require a few more definitions. 

\begin{definition}
    Define the \emph{transfer augmented} value function to be 
 \[ V^t(\mu) = \max_{a \in A} \left\{ \underbrace{\bb{E}_\mu[v(a, \theta) + k u(a, \theta)]}_{\text{\textcolor{budgreen}{Augmented Total Surplus}}} \right\} - \underbrace{k \bb{E}_\mu[u(a^*(\mu, \mathbf{0}), \theta)]}_{\textcolor{burgundy}{\text{Receiver Outside Option}}}. \]
\end{definition}

Implicitly, $V^t(\mu)$ specifies realized payments on path as exactly those which make Receiver indifferent between the surplus maximizing action and her own payoff maximizing action. It will turn out that this class of transfers are exactly those used at the optimum (see the proof of Proposition \ref{p: k-cavification}). 
Modulo transfers, I next need to characterize the beliefs supported by any optimal information policy. 

\begin{definition}
    Let $\mcal O_a = \{\mu : a \in a^\dagger(\mu, \mathbf{0})\}$. $\mu$ is \emph{extremal} if it is an extreme point of $\mcal O_a$. 
\end{definition}

Extremal beliefs are those which make the agent maximally indifferent, and have been shown to be useful in a variety of other distinct persuasion problems (see \cite{lipnowski_mathevet_2017} and \cite{gao2023priorfree}). In this setting, they are useful because they are the ``supporting'' points of the concavified value function: 

\begin{definition}
Fix a set $\mcal K \subset \Delta(\Theta)$. The \emph{$\mcal K$-cavification} of a function $f: \Delta(\Theta) \to \bb{R}$, denoted $\text{cav}^{\mcal K}(f)$, is the smallest concave function such that $\text{cav}^{\mcal K}(f)(\mu) \geq f(\mu)$ for all $\mu \in \mcal K$. 
\end{definition}

\begin{proposition}
    \label{p: k-cavification}
    There exists a finite set of extremal beliefs, $\mcal K$, such that $\text{cav}^{\mcal K}(V^t)(\mu) = V^*(\mu)$ for all $\mu \in \Delta(\Theta)$. 
\end{proposition}

Proposition \ref{p: k-cavification} is proven in \blue{\ref{p: k-cavification proof}} Note even though the transfer-augmented function has Receiver take surplus maximizing actions at every belief, Sender need not fully reveal the state, in contrast to standard intuition. This is because of the limited liability constraint: there may exist beliefs where Sender would prefer to charge Receiver if they could at the surplus maximizing action, but cannot. Consequently, Sender may do better by pooling that belief with one where Receiver is given less surplus (i.e. made maximally indifferent) to persuade Receiver to take actions which are more beneficial to Sender. Hence the ability to persuade can be strictly useful even when full information is ex-post efficient so long as Sender faces a limited liability constraint. 

Proposition \ref{p: k-cavification} simplifies the process of finding the transfer-augmented concavification in two ways. First, it shows that it is without loss to focus on a simple class of transfers, and gives an economic intuition for those transfers. Second, it shows that one need only focus on extremal beliefs when computing the optimal persuasion value. As $k \to \infty$, so that transfers are never used, this implies the following simple fact about finite persuasion models\footnote{So, for example, the famous Figure 1 of Kamenica and Gentzkow could not have been generated by a finite persuasion problem.}. 

\begin{corollary}
    \label{l: corollary}
For any finite $\Theta, A$, $\text{cav}(V_0(\mu))$ is piecewise affine. 
\end{corollary}

\subsection{The Patient Ergodic Case}
In general, optimal contracts can be difficult to explicitly characterize because the interplay between the (Markovian) dynamics and intermediate discount rates make the self-generating set intractable to compute outside special cases.\footnote{Indeed, \cite{LehrerShaiderman2025}, who study a related question under the stronger assumption Receiver is myopic, call characterizing the value of persuasion with a forward looking Receiver ``The grand question...[We do] not offer a solution for the optimal disclosure of information by a patient sender when the disclosure has future payoff implications. This question remains open and it seems challenging to tackle.''}
However, in some simple settings, significant progress can be made in characterizing the \textit{value} of the optimal policy, in line with the \cite{KamenicaGentzkow11} approach of finding the value of persuasion instead of the optimal policy. In this subsection, I give some general bounds on the value of persuasion in the patient case. 

Towards doing so, define the set 
\[ \Gamma(u, \mu) = \left\{(\gamma, m) \in \Delta(A \times \Theta) \times \bb{R}_+: \bb{E}_\gamma[u(a, \theta)] + m \geq u \text{ and } \gamma|_{\Theta} = \mu \right\} \]
to be all couplings over states and actions $\gamma$ and total payments $m$ where Receiver secures a payoff of at least $u$, and which marginalize on $\Theta$ into $\mu$.
Suppose moreover that $M$ is irreducible and aperiodic with ergodic distribution $\mu^\infty$. Recall 
\[ \und U(\mu^\infty) = (1 - \delta)\sum_{\xi = 0}^\infty \delta^\xi \bb{E}_{\mu^\infty}[u(a^*(\mu^\infty, \mathbf{0}), \theta)] \]
is the value of Receiver's outside option at the ergodic distribution $\mu^\infty$.
$\Gamma(\und U(\mu^\infty), \mu^\infty)$ plays an important role in bounding the value of persuasion in Proposition \ref{p: persuasion upper bound}. To operationalize the bound, I need one more definition. 

\begin{definition}
    $\sigma$ is \emph{stationary} if, for all $h^\xi, h^\zeta \in \supp(\bb{P}^\sigma)$, $\sigma(h^\xi) = \sigma(h^\zeta)$. 
\end{definition}

\begin{proposition}
    \label{p: persuasion upper bound} Let $M$ be irreducible and aperiodic with ergodic distribution $\mu^\infty$. Then 
\[ \lims_{\delta \to 1} \mcal V^*(\mu_0, \delta) \geq \sup_{(\gamma, m) \in \Gamma(\und U(\mu^\infty), \mu^\infty)} \left\{ \bb{E}_\gamma[v(a, \theta) - km] \right\} \]
Moreover if there exists $\bar \delta$ such that there is an optimal stationary $\sigma^*$ for all $\delta > \bar \delta$, then the inequality holds with equality. 
\end{proposition}
The formal proof can be found in the \blue{\ref{p: persuasion upper bound proof}} The intuition uses the exponential concentration of distributions to their steady state, regardless of the initial distribution to show that as $\delta \to 1$, ``essentially'' only Receiver individual rationality at the ergodic steady-state matters. Thus, if (and only if) Receiver prefers the induced joint distribution to their no-information payoff, then some stationary distribution leading to $\gamma$ is implementable. The upper bound can be seen as implying the individual rationality constraint characterizes the value of optimal \textit{stationary} contracts\footnote{With transfers, these are often those which trace out the efficient set when agents have relational incentives; see \cite{levin2003relational} and \cite{KolotilinLi2021}.} exactly. However, outside of special cases, knowing when there is an optimal stationary contract can be difficult. 

This result is related to the Receiver individual rationality constraint of \cite{RenaultSolanVieille13}, who characterize the equilibrium payoff set of repeated communication games. There are two important differences. First, my Sender has commitment and hence the set of feasible payoffs is larger, leading to a simpler and cleaner characterization of the best stationary upper bound (they require additional conditions to characterize exactly the payoff set). Second, the addition of transfers modifies Receiver constraints to allow for all joint couplings and transfer profiles where Receiver individual rationality hold. Perhaps surprisingly, this lower bound is independent of transfers insofar as they affect the details of Receiver incentive compatibility---transfers augment the payoff set exactly by allowing additional ``lump sum'' payments that relax Receiver's individual rationality constraint at the ergodic distribution. 

Bounding Sender's optimal payoff can also shed qualitative light on the nature of dynamic persuasion away from Receiver's individual rationality constraint. First, it implies that a sufficient condition for Sender to strictly benefit from dynamics is that Receiver's individual rationality constraint is not binding. 

\begin{corollary}
\label{cor: benefits from dynamics}
    Let $\gamma$ be the coupling concavifying $V^t(\mu_0)$. If $\bb{E}_\gamma[u(a, \theta)] > \und U(\mu^\infty)$, then $\lims_{\delta \to 1} \mcal V^*(\mu_0, \delta) > \text{cav}(V^t(\mu_0))$ unless $\text{cav}(V^t(\mu_0)) = \bb{E}_{\mu_0}[\max_{a \in A}[v(a, \theta)]]$. 
\end{corollary}

Corollary \ref{cor: benefits from dynamics} contrasts with Theorem 1 of \cite{ely2017beeps} and highlights the value of the obedience-based characterization in Proposition \ref{p: persuasion upper bound} over the recursive concavification approach. In particular, unlike \cite{ely2017beeps}, I can give easy to verify conditions under which Sender benefits from the ability to dynamically persuade; moreover these conditions are independent of the underlying Markov chain, and hence are easy to verify. 

Second, recall in a Markovian persuasion problem providing information to Receiver can have two effects: a backwards-looking dynamic incentivization effect, where information today is provided as a reward for taking Sender-favorable actions in past periods, and a forward-looking future informativeness effect, where information today is ``sticky'' and will affect the efficacy of persuasion tomorrow by shifting where Receiver's posterior belief will be. 
\cite{LehrerShaiderman2025} shut down the dynamic incentivization effect (by considering only myopic Receivers) and show the static persuasion payoff is an upper bound to Sender's value from dynamic persuasion, even as the discount rate tends towards $1$. In contrast, the following holds in the general model. Here, $\text{cav}(V^t)(\mu)$ is the static $\mcal K$-cavification from Proposition \ref{p: k-cavification}. 

\begin{corollary}
    \label{cor: lehrer shaiderman comparison}
    Let $M$ be irreducible and aperiodic with prior equal to the ergodic distribution $\mu_0 = \mu^\infty$. Then 
    \[ \lims_{\delta \to 1} \mcal V^*(\mu_0, \delta) \geq \text{cav}(V^t)(\mu_0).\]
\end{corollary}
\begin{proof}
    Let $(\hat \rho, \hat t, \hat \alpha)$ be the profile at the $\mcal K$-cavification. By Lemma \ref{l: indifference transfers are sufficient}, $\bb{E}_{\hat \rho}[\bb{E}_\mu[u(a^*(\mu, \mathbf{0}), \theta)]] = \bb{E}_{\hat \rho}\bb{E}_{\mu, \alpha}[[u(a, \theta) + t^I(\mu, a)]]$. But also $\bb{E}_{\hat \rho}[\bb{E}_\mu[u(a^*(\mu, \mathbf{0}), \theta)]] \geq \bb{E}_{\mu_0}[u(a^*(\mu_0, \mathbf{0}), \theta))]$ by Blackwell's theorem. Thus $(\gamma, m)$ induced by $(\hat \rho, \hat t, \hat \alpha)$ is in $\Gamma(\und U(\mu^\infty), \mu^\infty)$. Apply Prop. \ref{p: persuasion upper bound}.
\end{proof}

As $k \to \infty$, the $\mcal K$-cavification and concavification coincide. Thus, Corollary \ref{cor: lehrer shaiderman comparison} implies the value of static persuasion moves from an upper bound (with myopic Receiver) into a lower bound (with patient receiver). Economically, I see this result as implying the dynamic incentivization effect completely swamps the information effect, under exactly the assumptions in the Markovian persuasion literature. 
Corollary \ref{cor: lehrer shaiderman comparison} thus sheds light on the role that myopia plays in dynamic persuasion and serves as a cautionary tale against modeling myopic Receivers in applications where intertemporal tradeoffs are likely to be first order.

\section{Optimal Loyalty Contracts}

Consider the following stylized environment, inspired by the problem Uber must face when designing a dynamic contract. Suppose each period (i.e. a day) Uber (Sender) pushes $n$ rides to a driver (Receiver), the quality of which is unknown to the driver but privately observed by Uber. Suppose that Uber always wants the driver to accept the ride ($a_i = 1$) (though their value from acceptance may vary from ride to ride), so that the payoff from ride $i$ is given by $v_i(a_i, \theta_i) = c_i \mathbf{1}\{a_i = 1\}$ for some $c_i > 0$. The driver, however, wants to accept the ride if and only if the ride is good, $\theta_i = 1$; normalize their payoff so that $u_i(a_i, \theta_i) = \mathbf{1}\{\theta_i = a_i\}$.\footnote{In each dimension, this describes the judge-prosecutor game a la \cite{KamenicaGentzkow11} where Uber is the prosecutor. An alternative (less natural) interpretation of this setup is one where the prosecutor tries many simultaneous trials in front of a judge in each period, and can also pay the judge to make a decision (though as $k \to \infty$, I recover the case without payments).} 
To incentivize the driver to accept rides, Uber can use two tools: directly increase their payments, or engage in information design. 

Formally, represent the primitives in each problem as
\[ \{\Theta, A, v, u, \mu_0\} = \left\{ \prod_{i = 1}^n \Theta_i, \prod_{i = 1}^n A_i, \sum_{i = 1}^n v_i, \sum_{i = 1}^n u_i, \prod_{i = 1}^n \mu_0^i \right\} \]

This formulation encodes three assumptions. First, that payoffs across dimensions are additively separable. Second, that the efficiency of transfers $k$ (implicit in the model setup) is the same across dimensions. Third, that the driver's prior belief about the quality of the ride is independently drawn across periods according to some prior belief $\mu_0^i$ in dimension $i$, with $\mu_0 = \prod_i \mu_0^i$. Suppose moreover that $\theta_i \sim \mu_0^i$ across all time periods, so that ride $i$'s valuation is drawn i.i.d. across time, and that $\mu_0^i \in (0, \frac12)$ for each $i$, so that there is room for incentive provision in each dimension.  

What is the optimal contract in this setting? While Theorem \ref{t: transfers as last resort} gives a partial answer, it does not explicitly characterize an optimal contract. Luckily, in this simple case, it is possible to explicitly compute the Pareto frontier and consequently derive the optimal contract, which takes a (surprisingly) simple form. 
Throughout the remainder of this section, let $\rho_i^{BP}$  be the Bayesian persuasion optimal solution in each dimension, splitting beliefs into $\{0, \frac12\}$ with the appropriate Bayes plausible probabilities, and let $\rho_i^{FI}$ reveal the state. 

\begin{definition}
    A \emph{tiered loyalty contract} is a contract $\sigma$ and a sequence of increasing stopping times $\{T_i^*\}_{i = 1}^n$, $T_i^* \leq T_{i + 1}^*$ such that for $\bb{P}^\sigma$-almost all histories,
    \[ \rho_i(h^\xi; \sigma) = \begin{cases} \rho_i^{FI} \text{  if  } t > T_i^*(h^\xi)  \\ \rho_i^{BP} \text{   if  } t < T_i^*(h^\xi) \end{cases}. \]
\end{definition}

Tiered loyalty contracts allow for one period of indeterminacy ($T_i^*(h^\xi) = T$), which arises due to the integer programming problem when providing incentives. In general, though tiered loyalty programs are stark: they provide myopically optimal information up until a certain point at which the relationship matures and Sender's ``stock'' of \textit{future} informational incentives runs out. After this, Sender ``promotes'' the driver, gives them full information, and then pays them to take valuable rides.  
This transition happens ``dimension-by-dimension:'' Sender slowly unwinds their advantage by transitioning from the static optima directly to full information in each dimension separately; consequently, each transition $T_i^*$ can be thought of as a distinct ``tier'' in the relationship. 

The main result of this section is that tiered loyalty contracts are exactly optimal. 

\begin{theorem}
    \label{t: loyalty contracts}
   There is an optimal contract which features transfers as a last resort and is a tiered loyalty contract. 
\end{theorem}

Theorem \ref{t: loyalty contracts} is a deuteragonist of this paper. In particular, it explicitly solves for the dynamic optimal contract in a class of payoffs that generalize the leading example in the information design literature. Moreover, thus a contract is natural and has a ``bang-bang'' structure: myopically optimal information is provided until some random \emph{promotion time}, after which the state in each dimension is fully revealed. Importantly, even though the problem (and incentives) are separable, information is leveraged across dimensions---future information revelation in one dimension is used as an incentive tool to encourage the agent to take the high action in a different dimension. 

Theorem \ref{t: loyalty contracts} mirrors the loyalty contracts employed by Uber, who run a driver loyalty program with four tiers (``Blue,'' ``Gold,'' ``Platinum,'' and ``Diamond'') each of which have different benefits and differential access to information. I thus see my model as providing a potential explanation for Uber's driver loyalty program. 
Alternatively, with two projects (like the project selection example succeeding Proposition \ref{p: k-cavification} without a safe outside option), the time $\xi_i^*$ can be thought of as promotion: once the worker has chosen the ``bad-fit'' project enough times, the manager gives them information (and thus discretion) to choose projects which the employee would prefer, even if doing so runs against the manager's own interests. 

The proof can be found in \blue{\ref{t: loyalty contracts proof}} The approximate intuition is as follows. First, the Pareto frontier is governed by the summary statistic $\bb{P}(a_i = 1 | \theta_i = 0)$ at each time in each dimension, the rate at which persuasion is ``successful.'' For Sender to provide incentives via information, it must be that they commit to an information structure which decreases this probability, which gives Receiver a marginal unit of $1$ while costing Sender $-c_i$. So long as this probability is interior (between the static optimum and full information), the Pareto frontier is linear, and hence dynamic incentives can be ``backloaded'' using a similar squeezing procedure as Lemma \ref{l: squeezing lemma} to provide information in dimension $i$ tomorrow instead of today. Hence it is possible to provide incentives by promising full information in some dimension at some point in the future, until eventually the informational advantage is completely unwoven; at this point, Sender unwinds their entire informational advantage and transfers come in as a last resort to motivate the agent. 

\section{Discussion}

This paper studied a model of dynamic contracting where Sender can both pay and persuade as ways to motivate Receiver. Theorem \ref{t: transfers as last resort} characterizes the time structure of optimal transfers---there exists an optimal contract where payments occur on the path of play only if Sender has committed already to reveal the state forever afterwards. 
One interpretation of this result is that dynamic information endows Sender with a ``stock'' of informational incentives, which they can always draw down upon before turning to transfers. 
In simple settings---such as the Uber example---this observation implied to the optimality of tiered loyalty programs, where static persuasion remains optimal until the entire informational stock is exhausted, after which Sender fully reveals the state and simply pays Receiver. 

There are several natural extensions under which I conjecture the result continues to hold. A modified Theorem \ref{t: transfers as last resort} extends to any restricted set of experiments where there is a unique Blackwell maximal element. My backloading result may also extend to the case where transfers are nonlinear in cost but differentiable. The result should extend to cases where payoffs evolve over time and the state is noisily observed, as discussed after Theorem \ref{t: transfers as last resort}. 

Several questions about the model remain open and are fruitful areas for future research. First, information was free to generate in my model; I conjecture for suitably defined costs of information (such that full information is not infinitely costly), an adapted version of the result holds.  
Second, understanding the model with moral hazard seems like a natural enrichment of the model. 
Importantly, Lemma \ref{l: pullback lemma} may no longer apply, as the action is not perfectly monitored and thus the correct ``pullback utility'' at the exact optimum is no longer clear. Despite this, tools from the quota mechanisms literature could prove effective in characterizing approximately optimal contracts where payments are a last resort. 
Third, it is natural to consider the case with multiple Receivers, who are playing some type of coordination game against Sender (i.e. a dynamic version of \cite{mathevet2020information} and \cite{smolin2023information} with \cite{winter2004incentives} and \cite{halac2021rank}. 
Fourth, comparative statics are an open question. It remains to be seen how the stopping time $T_\sigma$ which determines when transfers are used in contracts where transfers are a last resort changes with model primitives, in particular the discount rate. As $\delta$ increases, intertemporal incentives become more attractive to the principal, though the value of future information also increases, so it is not obvious whether transfers get used sooner or later. 
Finally, characterizing explicitly dynamic contracts outside of the simple cases studied in the previous section remains an open question. 
Indeed, \cite{LehrerShaiderman2025} call the simpler version of this question, without transfers, the ``grand question'' of dynamic information design. 

\bibliography{biblio.bib}

\newpage 
\appendix
\section*{Appendix A: Omitted Proofs}
\makeatletter\def\@currentlabel{Appendix A.}\makeatother
\label{Appendix}

\subsection*{PROOF OF LEMMA \ref{l: dynamic revelation principle}}
\label{l: dynamic revelation principle proof}
\begin{proof}
    Fix an obedient contract $\sigma$. Construct $\sigma^*$ as follows: at any on-path history, replace (relabel) $s$ with the distribution of random signals $\{(\theta | s), a\}$ each with probability $\alpha(s)(a)$, i.e. recommend directly the induced posterior belief and also the recommended action. 
    Clearly this does not affect Receiver's information or their continuation incentives and hence the realized belief is $\theta | s$ and the recommended action is obeyed. 
    Second, note decreasing transfers as much as possible at actions not induced on path can only weaken incentive compatibility constraints and help support a conjectured equilibrium profile. 
\end{proof}

\subsection*{PROOF OF PROPOSITION \ref{p: FE SP equivalence}}
\label{p: FE SP equivalence proof}
\begin{proof}
    The first part is standard, noting that the maximizing set of the static problem is nonempty and that payoffs satisfy a transversality condition (by (BD)). 
    The second follows by setting $u'(\mu, a) = \mcal U(\{h^\xi, \mu, a\})$, so that the choice of future utility promise and payoffs today must be part of an optimum to the sequential formulation, as otherwise a profitable one-shot deviation exists in the sequential problem. 
\end{proof}

\subsection*{PROOF OF LEMMA \ref{l: slope of value at optimum}}
\label{l: slope of value at optimum proof}
\begin{proof}

The first statement. For any $(u, \mu_0)$ and $\{\rho, t, u', \alpha\} \in \mcal F(u, \mu_0)$, and $\varepsilon > 0$, let $\tilde V(u + \varepsilon, \mu_0)$ be the value for Sender by choosing at inputs $(u + \varepsilon, \mu_0)$ the (feasible) tuple $\{\rho, t, u' + \frac{\varepsilon}{\delta}, \alpha\}$. This tuple strictly increases Receiver utility to meet the higher utility promise without otherwise affecting any IC constraints, so in particular it must do weakly worse than the optimum: $V(u + \varepsilon, \mu_0) \geq \tilde V(u + \varepsilon, \mu_0)$. 

Suppose now the lemma is false, so $V_+'(u'(\mu, a), M\mu) > -k$ for some $u'(\mu, a)$ where $\mu \in \supp(\rho)$, $a \in \supp(\alpha)$. By construction, we have 
\begin{align*}
    V_+'(u, \mu_0) = \lims_{\varepsilon \to 0_+} \frac{V(u + \varepsilon, \mu_0) - V(u, \mu_0)}{\varepsilon} \geq \lims_{\varepsilon \to 0_+} \frac{\tilde V(u + \varepsilon, \mu_0) - V(u, \mu_0)}{\varepsilon}.
\end{align*}
Moreover, by construction, one has for any $\varepsilon > 0$ that 
\[ \tilde V(u + \varepsilon, \mu_0) - V(u, \mu_0) = \delta \bb{E}_{\rho, \alpha} \left[  V\left(u'(\mu, a) + \frac{\varepsilon}{\delta}, M\mu\right) - V(u'(\mu, a), M\mu) \right].
\]
Putting these two expressions together and utilizing the fact $V_+'(u'(\mu), \mu) > -k$ implies 
\begin{align*}
    V_+'(u, \mu_0) & \geq \lims_{\varepsilon \to 0_+} \frac{\tilde V(u + \varepsilon, \mu_0) - V(u, \mu_0)}{\varepsilon} 
     \\ & = \lims_{\varepsilon \to 0_+} \frac{\delta}{\varepsilon} \bb{E}_{\rho, \alpha} \left[  V\left(u'(\mu, a) + \frac{\varepsilon}{\delta}, M\mu\right) - V(u'(\mu, a), M\mu) \right] 
     \\ & = \bb{E}_{\rho, \alpha} \left[ V_+'(u'(\mu, a), M\mu) \right]> -k
\end{align*}
where the second equality interchanges the (finite) expectation and evaluates the right derivative explicitly. The final inequality uses the fact that $V_+'(\cdot, M\mu) \geq -k$ everywhere with strict inequality (by assumption) for at least one $(\mu, a) \in \supp(\rho, \alpha)$. 
This contradicts our assumption about $V_+'(u, \mu_0)$.

The second statement. Suppose not. Then we can find tuple $\{\rho, t, u', \alpha\} \in \mcal F(u, \mu_0)$ at some $(u, \mu_0)$  where $t(\mu, a) > 0$, $\mu \in \supp(\rho)$, $a \in \supp(\alpha)$, but $V_+'(u'(\mu, a), \mu) > -k$. 
Recall Receiver's incentive compatibility constraint when recommended $a$ at $\mu$ takes the form 
\[ (1 - \delta) \bb{E}_\mu[u(a, \theta) + t(\mu, a)] + \delta u'(\mu, a) \geq (1 - \delta)\bb{E}_\mu[u(a', \theta), \theta)] + \delta \und U(M\mu) \]
Consider now the alternative compensation scheme $\{\rho, \tilde t, \tilde u, a\}$ which leaves the information and action recommendations unchanged by changing the compensation scheme only at $\mu$: for some $\varepsilon > 0$ (restricted to be small enough that $\tilde t(\mu)$ remains positive), set $\tilde t(\mu, a) = t(\mu, a) - \varepsilon$, and set $\tilde u'(\mu, a) = u'(\mu, a) + \frac{1 - \delta}{\delta} \varepsilon$ at all other beliefs $\tilde t = t$ and $\tilde u' = u'$. Then by construction 
\[ (1 - \delta)\tilde t(\mu) + \delta \tilde u' = (1 - \delta)t(\mu) + \delta u' \]
which implies Receiver's incentive compatibility constraints are unperturbed under $\{\rho, \tilde t, \tilde u', a\}$ (as is their promise-keeping constraint). Hence, $\{\rho, \tilde t, \tilde u', a\}$ is feasible at $(u, \mu_0)$. Whenever belief $\mu$ is realized and $a$ is recommended, however, Sender's payoff differs by a value of 
\begin{align*}
    (1 - \delta)k(t(\mu, a) - \tilde t(\mu, a)) + \delta(V(\tilde u'(\mu, a), M\mu) - V(u'(\mu, a), M\mu)) 
    \\ = (1 - \delta)k\varepsilon + \left[ \delta V\left( u'(\mu, a) + \frac{1 - \delta}{\delta} \varepsilon, M\mu \right) - \delta V(u'(\mu, a), M\mu)\right].
\end{align*}
Dividing through by $\varepsilon$ and taking the limit as $\epsilon \to 0$ then yields that the change in Sender's payoff is given by 
\begin{align*}
    \lims_{\varepsilon \to 0_+} \left( (1 - \delta) k + \delta \frac{V\left( u'(\mu, a) + \frac{1 - \delta}{\delta} \varepsilon, M\mu \right) - V(u'(\mu, a), M\mu)}{\varepsilon} \right) 
    \\ = (1 - \delta) k + (1 - \delta) V_+'(u'(\mu, a), M\mu).
\end{align*}
Since $V_+'(u'(\mu, a), M\mu) > -k$ by assumption, there exists some $\varepsilon > 0$ at which this change in utility is strictly positive, and hence using $(\tilde t, \tilde u')$ instead of $(t, u')$ as payment at this history is a profitable deviation. This contradicts the assumption $\{\rho, t, u', \alpha\} \in \mcal F(u)$. 
\end{proof}

\subsection*{PROOF OF LEMMA \ref{l: pullback lemma}}
\label{l: pullback lemma proof}
\begin{proof}
Start from $u$-constrained optimal $\sigma$ and define $\bar \sigma$ as follows. 
Let $\rho^{FI}$ be the experiment that reveals the state, and suppose $\bar \rho(h^\xi; \bar \sigma) = \rho^{FI}$ always. 
Suppose Sender recommends some action for sure $\bar \alpha(\delta_\theta; \bar \sigma) = \delta_{a(\theta)}$, where $a(\theta) \in \argmax_{a \in A} u(a, \theta)$ is some action that maximizes Receiver's utility at that state, and which promises Receiver $u(\delta_\theta, a) = u^{RFI}(\delta_\theta)$ in the future at all recommended actions (and $\und U(\delta_\theta)$ otherwise). Suppose moreover that payments never occur on the path of play, $\bar t \equiv 0$. 
Clearly $\{\bar \rho, \bar t, \bar \alpha\}$ is a feasible contract at any $u \leq u^{RFI}(\mu_0)$ and gives Receiver a payoff of exactly $u^{RFI}(\mu_0)$. 

Now define the following contract $\sigma'$. Still always reveal the state at each history, and let $\alpha'$ be defined as follows: for each state $\theta$ and on-path $h^\xi$, set 
$\alpha'(h^\xi; \sigma')(\delta_\theta) = \bb{Q}_\xi^{\sigma}(\cdot | \theta) \in \Delta(A)$ 
and let transfers $t'$ be such that for each $a' \in \supp(\alpha'(h^\xi; \sigma')(\delta_\theta))$, 
\[ t'(h^\xi; \sigma')(\delta_\theta, a') = \max_{a \in A} u(a, \theta) - u(a', \theta).\]
This is the difference in payoffs between following the recommendation at $\bar \sigma$ and under $a'$. 
Note that at every history, Receiver's stage game payoff under $\{\bar \rho, t', \alpha'\}$ is given by 
\[ u(a', \theta) + t(h^\xi; \sigma') = \max_{a \in A} u(a, \theta) \]
given our choice of transfers. But this is their full-information payoff and hence they obtain $u^{RFI}(\mu_0)$; obedience of $\sigma'$ then follows from obedience of $\bar \sigma$, which follows since at each $a' \in \supp(\alpha'(h^\xi; \sigma')(\delta_\theta))$ and $\theta$, 
\[ (1 - \delta) \max_{a \in A} u(a, \theta) + \delta u^{RFI}(M\delta_\theta) \geq (1 - \delta) u(\tilde a, \theta) + \delta \und U(M\delta_\theta) \]
for any $\tilde a$ as the terms in $(1 - \delta)$ on the left are greater by definition and the term on the left against $\delta$ is greater by Blackwell's theorem. 

We thus now have a profile $\{\bar \rho, t', \alpha'\}$ that, by construction, satisfies $\bb{Q}_\xi^\sigma = \bb{Q}_\xi^{\sigma'}$ at every possible history and where the state is fully revealed, i.e. properties (1) and (2). Moreover, it gives (by construction) Sender the same expected payoff absent transfers. How much more must Sender pay Receiver in expectation under $\sigma'$? Note by construction additional payments are exactly equal to 
\[ (1 - \delta)\sum_{\xi = 0}^\infty \delta^\xi \bb{E}_{Q_\xi^{\sigma}} \left[ \max_{a \in A} u(a, \theta) - u(a', \theta) \right] = u^{RFI}(\mu_0) - \mcal U(\sigma) + \mcal T(\sigma), \]
where $\mcal T(\sigma)$ are normalized payments under $\sigma$. 
Here, we use the fact 
\[ (1 - \delta)\sum_{\xi = 0}^\infty \delta^\xi \bb{E}_{Q_\xi^\sigma}\left[\max_{a \in A} u(a, \theta)\right] = (1 - \delta)\sum_{\xi = 0}^\infty \delta^\xi \bb{E}_{\theta | \xi}\left[\max_{a \in A} u(a, \theta)\right] = (1 - \delta)\sum_{\xi = 0}^\infty \delta^\xi \bb{E}_{M^\xi\mu_0} \max_{a \in A} u(a, \theta) \]
because beliefs are a martingale so the ex-ante expected distribution of states at time $\xi$ is exactly $M^\xi\mu_0$. 
This implies that Sender's value is given by
\[ \mcal V(\sigma') = \mcal V(\sigma) - k\left(u^{RFI}(\mu_0) - \mcal U(\sigma) + \mcal T(\sigma) \right) \implies V(\sigma) - V(\sigma') \leq k(u^{RFI}(\mu_0) - \mcal U(\sigma)). \]
and thus implies $\sigma'$ satisfies the third property. 

Finally, the fourth property follows immediately by noting that because $V_+'(u, \mu_0)$ is in the steepest (and hence linear from the right) part of the Pareto frontier by assumption, any contract which transfers utility at rate at least $k$ (which is the third property) must be optimal. Taking $\sigma' = \sigma^*$ thus completes the proof. 
\end{proof}

\subsection*{PROOF OF LEMMA \ref{l: squeezing lemma}}
\label{l: squeezing lemma proof}
\begin{proof}
Let $\sigma^*$ be optimal and fix any $h^\zeta \in \mcal H^\zeta(\sigma^*)$. Modify $\sigma^*$ in the following way: at any history $h^\phi$ where $h^\zeta \succ h^\phi$ or $h^\phi$ and $h^\zeta$ are incomparable, do not change $\sigma^*$. If $t(h^\zeta)(\mu, a) = 0$ or $t(h^\zeta)(\mu, a) > 0$ but the state is revealed always afterwards, do not change $\sigma^*$. Finally, fix $(\mu, a)$ such that $t(h^\zeta)(\mu, a) > 0$ but there exists an experiment in the future which is not full information. Here we want to change $\sigma^*$: there are two cases. 

First, if $\mcal U(\{h^\zeta, \mu, a\} | \sigma^*) \geq u^{RFI}(\mu)$, then the utility promise already exceeds the maximum possible payoff form full-information. In this case, because by Lemma \ref{l: slope of value at optimum} 
\\ $V_+'(\mcal U(\{h^\zeta, \mu, a\} | \sigma^*)) = -k$, the logic of the pullback lemma implies there is a $\mcal U(\{h^\zeta, \mu, a\} | \sigma^*)$-constrained optimum, $\tilde \sigma$, which gives full information and pays Receiver an additional amount (over the potential payoff from the pullback) to meet the utility promise constraint which is also optimal. 
Modifying $\sigma^*$ so that $\sigma^* = \bar \sigma$ at any history which is not succeeded by $h^\zeta$ on the path of play, and otherwise replacing the continuation contract $\sigma^*(\cdot | h^\zeta)$ with $\bar \sigma(\cdot | h^\zeta) = \tilde \sigma$ delivers the desired continuation contract $\bar \sigma$ with $h^\zeta \not\in \mcal H^\zeta(\bar \sigma)$. 

Second, suppose $\eta = u^{RFI}(\mu) -\mcal U(\{h^\zeta, \mu, a\} | \sigma^*) > 0$. 
Take $\tilde u = \frac{1 - \delta}{\delta} t(h^\zeta; \sigma)(\mu, a)$. If $\eta > \tilde u$, then replace $t(h^\zeta)(\mu, a)$ with $0$, but add to $\mcal U(\{h^\zeta, \mu, a\})$ the amount $\tilde u$ exactly. Note Sender is indifferent to this change since $V_+e'(\mcal U(\{h^\zeta, \mu, a)) = -k$. The replacement procedure in the first case and the fourth property of the pullback lemma then imply there exists a sequentially optimal continuation contract starting from $h^\zeta$ at which payments are $0$ under $(\mu, a)$. 
If instead $\eta < \tilde u$, set $t(h^\zeta; \bar \sigma)(\mu, a) = t(h^\zeta; \sigma)(\mu, a) - \frac{\delta}{1 - \delta}\eta$ and set continuation contracts to be the obedient $\sigma^*$ in the pullback lemma. 
Because $\tilde u > \eta$, set $\alpha = 1$ in the fourth property of the pullback lemma, i.e. there is full revelation at all future continuation histories. 
As before, this maintains optimality of the continuation contract without affecting the joint distribution of actions in histories before $s$ (or incomparable to $h^\zeta$). 
In either case, transfers are now a last resort at $h^\xi$ conditional on $(\mu, a)$. Repeating this procedure for all $(\mu, a)$ that appear on path implies the argument. 
\end{proof}

\subsection*{PROOF OF PROPOSITION \ref{p: feasibly optimal actions}}
\label{p: feasibly optimal actions proof}
\begin{proof}
    Suppose not, so there exists $a \in \supp(\alpha(h^\xi)(\mu))$ occurring on path such that the condition fails. 
    Define $a'$ to be one such action where the following sequence holds: 
    \[ \bb{E}_\mu[v(a', \theta) - v(a, \theta)] \geq  \min_{\theta \in \Theta} v(a', \theta) - v(a, \theta) > k \max_{\theta \in \Theta} u(a, \theta) - u(a', \theta) \geq k \bb{E}_\mu[u(a, \theta) - u(a', \theta)].\]
    Let $\tilde t(\mu) = \bb{E}_\mu[u(a, \theta) - u(a', \theta)]$ be the additional payment necessary to engender $a'$ instead of $a$ at belief $\mu$. Note $\tilde t(\mu) \geq 0$ since we assumed we were at an optimum. 

    But now consider the alternative one-shot deviation to $\sigma$, occurring at $h^\xi$, where $a'$ is recommended (and payments are augmented by $\tilde t(\mu)$ at belief $\mu$), changing nothing else. By choice of payments, this is obedient for Receiver.
    This changes Sender's utility at $h^\xi$ by
    \[ \bb{E}_\mu[v(a', \theta) - v(a, \theta)] - k\bb{E}_\mu[u(a, \theta) - u(a', \theta)] > 0\]
    contradicting our assumption we were at optimum. 
\end{proof}

\subsection*{PROOF OF PROPOSITION \ref{p: strict optimality}}
\label{p: strict optimality proof}
\begin{proof}
It will be easier to prove the contrapositive of the statement, which is implied by the following result. 
\begin{lemma}
\label{l: belief effectiveness ratio}
    Fix interior belief $\tilde \mu$ with $E(\tilde \mu) > -k$ and optimal $\sigma$. If $\tilde \mu \in \supp(\rho(h^\zeta; \sigma))$ for on-path $h^\zeta$, then $t(h^\xi; \sigma)(\mu, a) = 0$ for all on-path $h^\xi$ where $h^\zeta \succ h^\xi$. 
\end{lemma}
\begin{proof}
    Suppose not, so $E(\tilde \mu) > -k$ but there exists an optimal $\sigma$ where $\tilde \mu \in \supp(\rho(h^\zeta; \sigma))$ but payments occur prior to $h^\zeta$. This implies there exists $h^\xi$ where $t(h^\xi; \sigma)(\mu, a) > 0$ for $a \in \supp(\alpha(h^\xi; \sigma)(\mu))$, $\mu \in \supp(\rho(h^\xi; \sigma))$, but $h^\zeta \succ (h^\xi, \mu, a)$.   
   Recall $\tilde \mu$ is nondegenerate;  consider an experiment $\tilde \rho$ which further reveals the state when $\tilde \mu$ would have been realized under $\rho(h^\zeta; \sigma)$, but otherwise is the same as $\rho(h^\zeta; \sigma)$; that is, 
    \[ \tilde \rho(\mu) = \begin{cases} \rho(h^\zeta; \sigma)(\mu) \text{  if  }  \mu \in \supp(\rho(h^\zeta; \sigma)) \setminus \{\tilde \mu\} \\ \tilde \mu(\theta) \text{  if  } \mu = \delta_\theta, \theta \in \supp(\tilde \mu) \\ 0 \text{  otherwise  } \end{cases}.  \]
    Here, we abuse notation to allow for $\tilde \rho(\mu)$ to ``duplicate'' two beliefs (if $\delta_\theta$ was, for example, already supported by $\rho(h^\zeta; \sigma)$). This is without loss of generality since Sender can recommend a mixed action at $\delta_\theta$, though the proof is clearer by allowing for these duplicate beliefs. 
    Suppose the action recommendation allows Receiver to take their favorite action at the split belief and otherwise recommends the same action: 
    \[ \tilde \alpha = \begin{cases} \alpha(h^\zeta; \sigma)(\mu)  \text{  if  }  \mu \in \supp(\rho(h^\zeta; \sigma)) \setminus \{\tilde \mu\} \\ \delta_{a^*(\mu, \mathbf{0})} \text{  if  } \mu = \delta_\theta, \theta \in \supp(\tilde \mu) \end{cases}. \] 
    That is, $\tilde \alpha$ lets Receiver take their favorite action when the state is revealed under $\tilde \rho$ but not $\rho(h^\zeta; \sigma)$ and otherwise does not change the action recommendation. 
    
   Moreover, consider payments $\tilde t$ defined by 
   \begin{align*}
       \tilde t(\mu, a) = \begin{cases} t(h^\zeta; \sigma)(\mu, a) \text{  if  }  \mu \in \supp(\rho(h^\zeta; \sigma)) \setminus \{\tilde \mu\}, a \in \supp(\tilde \alpha(h^\zeta; \sigma)(\mu)) \\ t(h^\zeta; \sigma)(\tilde \mu, a) \text{  if  } \mu = \delta_\theta, \theta \in \supp(\tilde \mu), a \in \supp(\tilde \alpha(h^\zeta; \sigma)(\mu))\end{cases} 
   \end{align*}
    and utility promises $\tilde u'$ defined as 
    \[ \tilde u'(\mu, a) = \begin{cases} u'(h^\zeta; \sigma)(\mu, a) \text{  if  }  \mu \in \supp(\rho(h^\zeta; \sigma)) \setminus \{\tilde \mu\}, a \in \supp(\tilde \alpha(h^\zeta; \sigma)(\mu)) \\ u'(h^\zeta; \sigma)(\tilde \mu, a) \text{  if  } \mu = \delta_\theta, \theta \in \supp(\tilde \mu), a \in \supp(\tilde \alpha(h^\zeta; \sigma)(\mu)) \end{cases}.\]
   Note that $(\tilde \rho, \tilde t, \tilde u', \tilde \alpha)$ maintain both the same expected transfers and the same path of utility promises as $(\rho(h^\zeta; \sigma), t(h^\zeta; \sigma), u'(h^\zeta; \sigma), \alpha(h^\zeta; \sigma))$ and thus have no effect on payments or Sender's continuation value after $h^\zeta$. 

    Together, this gives a tuple $\{\tilde \rho, \tilde t, \tilde u', \tilde \alpha\}$ which increases Receiver's utility by 
    \[ \eta = \bb{E}_{\tilde \mu}[u(a^*(\delta_\theta, \mathbf{0}), \theta)) - u(\alpha(h^\zeta; \sigma)(\tilde \mu), \theta)] \geq \bb{E}_{\tilde \mu}[u(a^*(\delta_\theta, \mathbf{0}), \theta)) - u(a^*(\tilde \mu, \mathbf{0}), \theta)] > 0. \]
    The first inequality follows from the definition of $a^*(\tilde \mu, \mathbf{0})$, and the second as full information is strictly optimal for Receiver. 
    This decreases Sender's utility at history $h^\zeta$ by 
    \[  \varepsilon = \bb{E}_{\tilde \mu}[v(a^*(\delta_\theta, \theta)) - v(\alpha(h^\zeta; \sigma)(\tilde \mu), \theta)] \geq  \min_{a' \in \mathscr{F}} \{ \bb{E}_{\tilde \mu}[v(a^*(\delta_\theta, \theta)) - v(a', \theta)] \}. \]
    Note here that $\text{supp}(\alpha)(h^\zeta; \sigma)(\tilde \mu) \in \mathscr{F}$ by Proposition \ref{p: feasibly optimal actions}.  
Together, this implies $\frac{\varepsilon}{\eta} > -k$. If $\epsilon \geq 0$ then this is obvious. Otherwise, if $\epsilon < 0$, then 
\begin{align*}
    \frac{\varepsilon}{\eta} 
& \geq \frac{\varepsilon}{\bb{E}_{\tilde \mu}[u(a^*(\delta_\theta, \mathbf{0}), \theta)) - u(a^*(\tilde \mu, \mathbf{0}), \theta)]}
\\ & \geq \min_{a' \in \mathscr{F}} \frac{\bb{E}_{\tilde \mu}[v(a^*(\delta_\theta, \theta)) - v(a', \theta)]}{\bb{E}_{\tilde \mu}[u(a^*(\delta_\theta, \mathbf{0}), \theta)) - u(a^*(\tilde \mu, \mathbf{0}), \theta)]} = E(\tilde \mu) > -k. 
\end{align*}
The first inequality decreases the denominator of a (negative) fraction; the second decreases the value of the numerator; the third by the definition of $E(\tilde \mu)$, and the fourth by assumption. 

From here, fix some $\beta \in (0, 1)$ be chosen so that $\beta \delta^{\zeta - \xi} \eta \bb{P}^\sigma(h^\zeta | (h^\xi, \mu)) \leq t(h^\xi; \sigma)(\mu)$.
Consider the alternative contract $\tilde \sigma$, which modifies $\sigma$ in the following way: 
\begin{enumerate}
    \item For any history $h^\phi$, $h^\phi \not\succsim h^\xi$, $\tilde \sigma = \sigma$. 
    \item At $h^\xi$, $t(h^\xi; \tilde \sigma)(\mu, a) = t(h^\xi; \sigma)(\mu, a) - \beta \delta^{\zeta - \xi} \eta \bb{P}^\sigma(h^\zeta | (h^\xi, \mu))$. At any continuation contract $h^\phi \succ h^\xi$ where $h^\phi \not\succsim h^\zeta$, do not change the contract from $\sigma$. 
    \item At $h^\zeta$, with probability $\beta$, set $\sigma^S(h^\zeta) = \{\tilde \rho, \tilde t\}$ with continuation payoffs $\tilde u'$, and set $\sigma^R(h^\zeta) = \tilde \alpha(h^\zeta)$. With complementary probability $1 - \beta$, do not change the continuation contract; that is, $\sigma^S(\cdot | h^\zeta) = \tilde \sigma^S(\cdot | h^\zeta)$.  
    \item When the probability $\beta$ event in Step (3) occurs, at every belief $\mu \in \supp(\tilde \rho)$, choose $\tilde \sigma(\cdot | (h^\zeta, \mu))$ to be some $\tilde u'$-constrained optimal continuation contract. 
\end{enumerate}
Note that by choice of $\eta$ and construction of $\tilde \sigma$, Receiver incentive compatibility constraints remain unchanged at $h^\xi$ and thus their actions only change at $h^\zeta$. Thus $\tilde \sigma$ is feasible. 

This observation then implies Sender's payoff changes only at two histories: $h^\xi$ and $h^\zeta$. At $h^\xi$, they pay $k \beta \delta^{\zeta - \xi} \eta \bb{P}^\sigma(h^\zeta | (h^\xi, \mu))$ less in transfers, while at $h^\zeta$ they forego $\beta \epsilon$ total payoff in order to do so. This changes their total expected payoff by 
\[ \delta^\xi \bb{P}^{\tilde \sigma}(h^\xi) \left(k \beta \delta^{\zeta - \xi} \eta \bb{P}^\sigma(h^\zeta | (h^\xi, \mu)) + \beta \delta^{\zeta - \xi} \varepsilon \bb{P}^{\tilde \sigma}(h^\zeta | (h^\xi, \mu))\right). \]
where we note $\bb{P}^\sigma(h^\zeta | (h^\xi, \mu)) = \bb{P}^{\tilde \sigma}(h^\zeta | (h^\xi, \mu))$. This expression is strictly positive if and only if $k > \frac{-\varepsilon}{\eta}$, which is true by the way we have constructed $(\varepsilon, \eta)$. Hence $\tilde \sigma$ must give Sender a strictly higher payoff than $\sigma$, contradicting our assumption $\sigma$ was optimal. 
\end{proof}

It is clear that Lemma \ref{l: belief effectiveness ratio} implies Proposition \ref{p: strict optimality}. 
\end{proof}

\subsection*{PROOF OF PROPOSITION \ref{p: Ek convex}}
\label{p: Ek convex proof}
\begin{proof}
    The second statement is obvious; if $k' < k$, then $E(\mu) > -k'$ implies $E(\mu) > -k$.
    The first statement. Fix $\mu, \mu' \in \mcal E(k)$. Note that 
    \[ \und v(\mu) = \min_{a' \in \scr{F}} \bb{E}_\mu[v(a^*(\delta_\theta, \mathbf{0}), \theta) - v(a', \theta)] \]
    is the minimum of linear functions in $\mu$ and hence concave. Moreover, we know that 
    \[ \und u(\mu) = \min_{a \in A} \bb{E}_\mu[u(a^*(\delta_\theta, \mathbf{0}), \theta) - u(a, \theta)] = \bb{E}_\mu[u(a^*(\delta_\theta, \mathbf{0}), \theta) - u(a^*(\mu, \mathbf{0}, \theta))] \]
    is also concave in $\mu$. Setting $\beta \mu + (1 - \beta)\mu' = \mu \beta \mu'$, we have 
    \[ \frac{\und v(\mu \beta \mu')}{\und u(\mu \beta \mu')} > -k \iff \und v(\mu \beta \mu') > -k \und u(\mu \beta \mu').  \]
   Because $k > 0$ and $\und u(\cdot)$ is concave, we have that 
   \[  -k \beta \und u(\mu) - k(1 - \beta)\und u(\mu') \geq -k \und u(\mu \beta \mu'). \]
But we also know $\und v(\mu \beta \mu') \geq \beta \und v(\mu) + (1 - \beta) \und v(\mu')$ by concavity of $\und v$. Finally, since $\mu, \mu' \in \mcal E(k)$, $\und v(\mu) > -k \und u(\mu)$ and $\und v(\mu') > -k \und u(\mu')$. Putting everything together, we have that 
\[ \und v(\mu \beta \mu') \geq \beta \und v(\mu) + (1 - \beta) \und v(\mu') > -k \beta \und u(\mu) - k (1 - \beta) \und u(\mu') \geq -k \und u(\mu \beta \mu') \]
which gives the desired argument.
\end{proof}

\subsection*{PROOF OF PROPOSITION \ref{p: weak ray regularity}}
\label{p: weak ray regularity proof}
\begin{proof}
The forward direction. 
    Suppose primitives are nontrivial and incentivizable at $\mu_0$ and suppose transfers are not strictly backloaded.
    Then we can find an optimal contract $\{\rho, t, u', \alpha\} \in \mcal F(0, \mu_0)$ such that $u'$ is identically $0$. Yet payoffs are incentivizable, and so there is some $\mu \in \supp(\rho)$ where the incentive compatibility constraint written at $a \in \supp(\alpha(\mu))$ on path can be written as 
    \[ (1 - \delta) \bb{E}_\mu[u(a, \theta) + t(\mu, a)] + \delta u'(\mu, a) \geq (1 - \delta) \bb{E}_\mu[u(a^*(\mu, \mathbf{0}), \theta)]\]
    where $a \neq a^*(\mu, 0)$. 
  Since $u'(\mu, a) = 0$, it must be that $t(\mu, a) > 0$. 
  Consider now the optimal contract $\{\tilde \rho, \tilde t, \tilde u', \tilde \alpha\}$ differing from $\{\rho, t, u', \alpha\}$ only at $(\mu, a)$ where $\tilde t(\mu, a) = t(\mu, a) - \varepsilon$ but $u'(\mu, a) = \frac{1 - \delta}{\delta} \varepsilon$. Incentive constraints are unperturbed, and hence this increases Sender's value by $(1 - \delta) k\varepsilon$ (in savings of transfers), at a cost of $\delta(V(0) - V(\frac{1 - \delta}{\delta} \varepsilon))$.  
  Because the problem is nontrivial, it is clear that there is $\varepsilon > 0$ small enough where this alternative contract is strictly profitable, contradicting optimality of the original contract. 

  The converse. Suppose first that the problem was not incentivizable. Then there is an optimal contract where for each $(\mu, a)$ on path, $a = a^*(\mu, \mathbf{0})$. Then this is clearly still an optimal contract when $u'(\mu, a) = 0$ always (promise keeping at $0$ never binds, and incentive compatibility is trivial). Hence transfers cannot be strictly backloaded. 

  Suppose instead that the problem was trivial. Fix any optimal contract $\{\rho, t, u', \alpha\} \in \mcal F(0, \mu_0)$ where $u'(\mu, a) > 0$. Because the problem is trivial, the contract $\{\rho, t + \frac{\delta}{1 - \delta} u', 0, \alpha\}$ must also be optimal because it simply shifts utility payments from tomorrow to today at a constant marginal cost. Yet this contract is one that ensures transfers are not strictly backloaded.     
\end{proof}

\subsection*{PROOF OF PROPOSITION \ref{p: wrr if static k-cavification}}
\label{p: wrr if static k-cavification proof}
\begin{proof}
      Suppose incentivizability does not hold. Then there is an optimal contract $\{\rho, t, u', \alpha\}$ where $a = a^*(\mu, \mathbf{0})$ in the first period. Since the utility promise is $0$ in the first period, this in particular implies that there exists an optimal contract of the form $\{\rho, 0, 0, \alpha\}$ because incentive constraints never bind and $V$ is nonincreasing. 
    But this contract must be optimal among all contracts under which $t, u' \equiv 0$, so in particular it gives the value of $\text{cav}(V_0)(\mu_0)$. It must also be optimal among all \textit{stationary} contracts where $t > 0$ is allowed, $u' = 0$, and hence it gives a value of $\mcal V^*(\mu_0, 0)$. This implies that transfers are not useful in the static problem, completing the contrapositive. 
\end{proof}

\subsection*{PROOF OF PROPOSITION \ref{p: always last resort}}
\label{p: always last resort proof}
\begin{proof}
    Define first the value function $V^0$ to be the value function obtained from (FE) with the additional restriction that transfers must always be 0, i.e. there are no transfers. This is the pure dynamic information design problem. Note the domain of $V(\cdot, \mu_0, \delta)$ is $[0, u^{RFI}(\mu_0)]$. Define $-\bar k = \inf_{\mu_0 \in \Delta(\Theta)} V_-^{0'}(u^{RFI}(\mu_0))$ to be a lower bound on the left derivative of $V^0$ at its right endpoint $u^{RFI}(\mu_0)$. 
    Suppose now the result is false, so there exists $k > \bar k$ and optimal $\sigma$ at primitives $\{u, v, \delta, k\}$ such that transfers are not a last resort. This implies there exists $h^\zeta \succ (h^\xi, \mu, a)$ on the path of play such that $\rho(h^\zeta; \sigma) \neq \rho^{FI}$ where $t(h^\xi; \sigma)(\mu, a) > 0$. 
    
    Let $\mu^\zeta$ be the belief Receiver has history $h^\zeta$. By mechanically pushing transfers back into history $h^\zeta$, the fact $t(h^\xi; \sigma)(\mu, a) > 0$ implies there exists solution $\{\rho, t, u', \alpha\} \in \mcal F(\mcal U(h^\zeta; \sigma), \mu^\zeta)$ where $\rho \not\in \rho^{FI}$ and $t \neq 0$ everywhere. Yet because transfers were used at $h^\xi$ preceding $h^\zeta$, $V_+'(\mcal U(h^\zeta; \sigma)) = -k$ by Lemma \ref{l: slope of value at optimum}. Yet this contradicts the fact $\rho \neq \rho^{FI}$, since $V_+^0(\cdot, \mu^\zeta) \geq -\bar k > -k$, so just providing informational incentives must be more efficient than $-k$. In particular, this implies decreasing transfers $t$ towards $0$, and increasing $\rho$ towards $\rho^{FI}$ must be a profitable improvement over $\{\rho, t, u', \alpha\}$, contradicting the assumption that tuple was optimal. 
\end{proof}

\subsection*{PROOF OF PROPOSITION \ref{p: k-cavification}}
\label{p: k-cavification proof}
\begin{proof}
    The proof follows two steps. First, we show that it is sufficient to focus on \textit{canonical} transfers, those defined by 
\[ t^I(\mu, a) = \bb{E}_\mu[u(a^*(\mu, \mathbf{0}), \theta) - u(a, \theta)] \mathbf{1}\left\{a \in a^S(\mu)\right\} \]
where 
\[ a^S(\mu) = \argmax_{a \in A} \bb{E}_\mu[v(a, \theta) - k(u(a^*(\mu, \mathbf{0}), \theta) - u(a, \theta))].\] 

\begin{lemma}
    \label{l: indifference transfers are sufficient}
    For any prior $\mu_0$ and optimal $(\rho^*, t^*)$, the tuple $(\rho^*, t^I)$ is optimal as well. 
\end{lemma}
\begin{proof}
    Fix an optimal $(\rho^*, t^*)$, and suppose $a^*(\mu, t^*)$ is induced at some belief $\mu \in \supp(\rho^*)$ inducing $a^*(\mu, t^*)$. First suppose $a^*(\mu, t^*) \in a^S(\mu)$. Since Receiver takes action $a^*$, this implies
    \[ \bb{E}_\mu[u(a^*(\mu, t^*), \theta)] + t^*(\mu, a^*(\mu, t^*), ) \geq \bb{E}_\mu[u(a, \theta)] + t^*(\mu, a) \text{  for all  } a \in A \]
    Taking differences implies  
    \[ \bb{E}_\mu[u(a^*(\mu, t^*), \theta) - u(a, \theta)] \geq t^*(\mu, a^*(\mu, t^*)) - t^*(\mu, a) \text{  for all   } a \in A \]
   Since $a^*(\mu, \mathbf{0})$ maximizes Receiver's payoff without transfers, we have that for any $a \not\in a^S(\mu)$ (so that $t^I(\mu, a) = 0$)
   \begin{align*}
        \bb{E}_\mu[u(a^*(\mu, t^*), \theta) - u(a, \theta)] \geq  \bb{E}_\mu[u(a^*(\mu, t^*), \theta) - u(a^*(\mu, \mathbf{0})), \theta)] = t^I(\mu, a^*(\mu, t^I))
   \end{align*}
and so in particular under payments $t^I$ Receiver will not want to take any $a \not\in a^S(\mu)$. 
Suppose now that $a^*(\mu, t^*) \not\in a^S(\mu)$. Then if at belief $\mu$ Sender paid $t^I(\mu, a)$ for some $a \in a^S(\mu)$, they could attain a strictly higher payoff then the induced pair under $(\rho^*, t^*)$ at $\mu$, a contradiction to the optimality of the original tuple. This finishes the proof. 
\end{proof}

Given that canonical transfers are sufficient, it is clear that 
\[  \max_t \bb{E}_\mu[v(a^*(\mu, t), \theta) - k t(\mu, a))]   = \max_{a \in A} \bb{E}_\mu[v(a, \theta) + k(u(a, \theta) - u(a^*(\mu, \mathbf{0}), \theta))]\]
and hence $V^*(\mu_0)= \text{cav}|_{\mu_0}(V^t(\mu))$ always. The next step shows that this concavification is equal to the k-cavification on extremal beliefs. 

To do so, we need to prove a quick geometric lemma.
\begin{lemma}
    \label{O_a properties}
    $\mcal O_a$ is a convex polytope. 
\end{lemma} 
\begin{proof}
    $\mcal O_a$ is compact and convex by Lemma A.1 of \cite{gao2023priorfree}; that $\mcal O_a$ is a polytope follows by noting 
\[ \mcal O_a = \Delta(\Theta) \cap \left( \A_{a' \in A \setminus \{a\}} \left\{ m: \int u(a, \theta) - u(a', \theta) dm \geq 0 \right\} \right)  \] $\mcal K = \U_a \text{ext}(\mcal O_a)$. 
where $m$ is any measure (not necessarily a probability measure). This is a finite intersection of half-spaces and thus each $\mcal O_a$ has finitely many extreme points by Theorem 19.1 of \cite{convex_rockafellar}. 
\end{proof}

We can now prove the final step. For any fixed $\tilde a$, recall that on $\mcal O_{\tilde a}$,  
\[ V^t(\mu) = \max_{a \in A} \left\{ \bb{E}_\mu[v(a, \theta) - k (u(\tilde a, \theta) - u(a, \theta))] \right\}, \]
which is the maximum of linear functions over a finite index.
This implies $V^t(\mu)$ is convex over the interior of each $O_{\tilde a}$. 
Moreover, $V^t(\mu)$ is globally upper semi-continuous over all of $\mu$ since it is the finite upper envelope of continuous functions. Thus, we have that $\lims_{\mu \to \bar \mu} V^t(\mu) \leq V^t(\bar \mu)$ for any $\bar \mu \in \Delta(\Theta)$, with strict inequality only potentially possible on the interior of $O_a$. 

Now set $\mcal K = \U_a \text{ext} O_a$ be the set of extremal beliefs; this is finite. By definition, it must be that $\text{cav}^{\mcal K}(V^t)(\mu)$ is a concave function such that $\text{cav}^{\mcal K}(V^t)(\mu) \geq V^t(\mu)$ for all $\mu \in \mcal K$; moreover $V^t$ itself is piecewise convex and globally upper semi-continuous (with jumps at most on points in $\mcal K$). 
But then because $\text{cav}^{\mcal K}(V^t)(\mu) \geq V^t(\mu)$ on the boundary of each $\mcal O_a$ and affine on the interior (by the definition of the concavification), it must be that  $\text{cav}^{\mcal K}(V^t)(\mu) \geq V^t(\mu)$ for each belief $\mu \in \Delta(\Theta)$. Since we know also $\text{cav}^{\mcal K}(V^t)(\mu) \leq V^*(\mu)$ for all $\mu \in \Delta(\Theta)$ it must be that $\text{cav}^{\mcal K}(V^t)(\mu) = \text{cav}(V^t)(\mu)$. 
 This finishes the proof. 
\end{proof}

\subsection*{PROOF OF PROPOSITION \ref{p: persuasion upper bound}}
\label{p: persuasion upper bound proof}
\begin{proof}
The lower bound. 
Define $\Gamma^{int}(\mu_0) = \{(\gamma, m) : \bb{E}_\gamma[u(a, \theta)] + m > \und U(\mu_0) \text{  and  } \gamma|_{\Theta} = \mu_0\}$. 
We show for any $\varepsilon > 0$ there exists $(\gamma, m) \in \Gamma^{int}(\und U(\mu^\infty))$ and $\bar \delta < 1$ such that there is some obedient $\sigma^*$ which gives Sender a payoff of at least 
$\bb{E}_\gamma[v(a, \theta) - km - \varepsilon]$ for all $\delta > \bar \delta$. 
Since payoffs are continuous over the (closed) set $\Gamma(u)$, this then implies the result. 

Fix some $(\gamma, m) \in \Gamma^{int}(\mu^\infty)$. 
Let $\sigma^*$ be the stationary contract where $\rho$ reveals the state and recommends mixed action $\gamma_\theta = \bb{E}[\gamma | \theta] \in \Delta(A)$ at $\delta_\theta$ at every history on-path, and pays $m$. If Receiver deviates from any recommendation, then give no information forevermore and recommend Receiver's no-information optimum.
At any history, any state $\theta$, and any action $a$, Receiver's deviation gain is bounded from above by 
\[ B = \max_{a', \theta'} u(a', \theta') - \min_{a, \theta} u(a, \theta) \] 
today. Receiver's incentive compatibility requirement can be written at $\mu = \delta_\theta$ and recommendation $a$ as
\[ (1 - \delta)( u(a, \theta) + m) + \delta \left( \bb{E}_\gamma[u(a, \theta)] + m \right) \geq (1 - \delta) u(a^*(\delta_\theta, \mathbf{0}), \theta) + \sum_{\xi = 1}^\infty \delta^\xi u(a^*(M^\xi\delta_\theta, \mathbf{0}), \theta). \]
Recall now the following theorem about the mixing times of Markov chains:  
\begin{proposition}
    [Levin and Peres, Theorem 4.9] 
    \label{Markov chain mixing} Let $M$ be irreducible and aperiodic. There exist $C, \alpha > 0$ such that $\max_{\theta \in \Theta} ||M^\xi \delta_\theta - \mu^\infty|| \leq C \alpha^\xi$ for $C, \alpha \geq 0$. 
\end{proposition}

Proposition \ref{Markov chain mixing} implies that for all $\xi$ and any $\eta > 0$, there exists $\bar T$ such that (1) for any $\xi \geq \bar T$, 
$||\bb{Q}_\xi^{\sigma^*}(h^\zeta | h^\xi) - \gamma||_\infty < \eta$ and (2) $|u(a^*(M^\xi \delta_\theta, \mathbf{0}), \theta) - u(a^*(\mu^\infty, \mathbf{0}), \theta)| < \eta$. Hence given any history, both players believe play to be close to $\gamma$ and that the distribution of the state will be close to $\mu^\infty$. 
This implies the value of deviating and receiving no information can be bounded above by 
\[ (1 - \delta) \sum_{\xi = 0}^{\bar T} B + \delta^{\bar T} \left( \und U(\mu^\infty) - \bb{E}_\gamma[u(a, \theta) + m] - \frac12 \varepsilon(\eta) \right) \]
where $\varepsilon(\eta)$ is a function vanishing as $\eta \to 0$. This holds regardless of the realized state. This then implies we can find $\bar \delta(\eta) < 1$ such that whenever $\delta > \bar \delta(\eta)$, the value of deviating is no more than $\varepsilon(\eta)$. Setting payments $m = \varepsilon(\eta)$ and taking $\eta \to 0$ then implies $\sigma^*$ is obedient. 

Finally, since $||\bb{Q}_\xi^{\sigma^*} - \gamma||_\infty$ vanishes uniformly as $\xi \to \infty$, Sender attains a payoff arbitrarily close to $\gamma$ as $\xi \to \infty$. This implies the lower bound (taking $\bar \delta$ big enough). 

The upper bound. Suppose there is an optimal stationary $\sigma^*$, so that there is a fixed $\{\rho, t, \alpha\}$ realized after every on-path history. By Proposition \ref{Markov chain mixing}, there exists $\bar T$ sufficiently large such that for all $\xi \geq \bar T$, $||Q_\xi^{\sigma^*} - \gamma||_\infty < \varepsilon$, where $\gamma$ is the coupling induced by $\sigma^*$ at the ergodic distribution.
Let $m$ be the induced total transfers at $\gamma$. 
The argument for the lower bound implies $\bb{E}_\gamma[u(a, \theta)] + m > \und U(\mu^\infty)$, as otherwise Receiver can deviate and secure at least their no-information payoff essentially always as $\delta \to 1$.
This also implies Sender cannot do better in this stationary equilibrium: in particular, for every $\varepsilon > 0$, there exists $\hat T$, independent of $\delta$, such that their payoff is bounded from above by 
\[(1 - \delta) \sum_{\xi = 0}^{\hat T} \delta^\xi \max_{a, \theta} v(a, \theta) + \delta^{\hat T} \left( \bb{E}_\gamma[v(a, \theta)] - km + \varepsilon\right). \]
Taking $\delta \to 1$ implies the upper bound and hence the proof. 
\end{proof}

\subsection*{PROOF OF THEOREM \ref{t: loyalty contracts}}
\label{t: loyalty contracts proof}
\begin{proof}
    The proof proceeds in four steps. As a preamble, we note that Sender never benefits from intratemporally coordinating information about rides in any given period because decisions are simultaneous and independent across dimensions. The first step is to characterize the way incentives are provided along the Pareto frontier.
    \begin{lemma}
    \label{l: example optimal}
        Let $\sigma$ be optimal. Then for any $h^\xi \in \supp(\bb{P}^\sigma)$, $\bb{P}(a_i = 0 | \theta_i = 1, h^\xi, \sigma) = 0$. 
\end{lemma}
\begin{proof}
Suppose not. Then there exists on-path history $h^\xi$ where $\bb{P}(a_i = 0 | \theta_i = 1, h^\xi, \sigma) > 0$. This implies there is a belief $\tilde \mu_i \in \supp(\rho_i(h^\xi; \sigma))$ such that $a_i(\tilde \mu) = 0$ but $\tilde \mu_i \neq \delta_0$.  

Consider the experiment $\tilde \rho_i$ which splits beliefs along $\tilde \mu_i$ in the following way. First, for any $\mu_i \neq \tilde \mu_i$, let $\rho_i(\mu_i) = \tilde \rho_i(\mu_i)$. Along $\tilde \mu_i$, split beliefs so that $\tilde \rho_i(\tilde \mu_i) = 0$, and $\tilde \rho_i(\delta_0) = \rho_i(\tilde \mu_i)\tilde \mu_i(0)$, and $\tilde \rho_i(\delta_1) = \rho_i(\tilde \mu_i)\tilde \mu_i(1)$. This is clearly Bayes plausible. 
In a similar way, modify payments and utility promises so that they are the same at beliefs where $\rho_i = \tilde \rho_i$, and equal to $t_i(\tilde \mu_i)$ and $u_i'(\tilde \mu_i)$ at the newly induced beliefs. 
    Finally, allow Receiver to take her favorite action at the newly induced beliefs, and otherwise the same action as before. 
    
    Such an experiment must be incentive compatible because Receiver is myopically maximizing at the new beliefs, and at the old ones (i.e. those supported by $\rho$), they are the same. Do not change any $\rho_j$, $j \neq i$.
    This alternative contract strictly increases $\bb{P}(a_1 = 1 | \theta_1 = 1, h^\xi, \sigma)$ while decreasing $\bb{P}(a_1 = 0 | \theta_1 = 1, h^\xi, \sigma)$ without affecting decisions in any other dimension. This increases Sender's stage-game payoff. Moreover, it does not change Sender's expected payments or continuation value (as the distribution of $\tilde t$ and $\tilde u'$ are the same as $t$ and $u'$). Thus this deviation is strictly profitable, contradicting supposed optimality of $\sigma$. 
\end{proof}

Lemma \ref{l: example optimal} implies that every optimal experiment can only induce action $a_i = 0$ at belief $\delta_0$; pooling all beliefs at which $a_i = 1$ is recommended then implies that the optimal information takes the form of $\supp(\rho_i) = \{0, \mu_i\}$, with $a_i = 1$ at $\mu_i$. The next step is to bound the cost of providing incentives by increasing $\mu_i$ in some dimension $i$. For the next lemma, let $V_i(u)$ be the value function of the problem in dimension $i$ only, and let $u_i^{RFI}$ be the value of full information to Receiver in only dimension $i$. 

\begin{lemma}
\label{l: slope of the pareto frontier}
     We have $V_{i+}'(u) = - c_i \text{  if  } u \leq u^{RFI}(\mu_0), c_i < k$, and $V_{i+}'(u) = -k$ otherwise. 
\end{lemma}
\begin{proof}
    Note that by Lemma \ref{l: example optimal}, it must be that the only way to provide incentives via information is by increasing $\mu_i$. This linearly decreases $\bb{P}(a_1 = 1 | \theta_1 = 0)$, benefiting Receiver by $1$ while costing Sender $-c_i$. 
\end{proof}

Lemma \ref{l: slope of the pareto frontier} and separability across dimensions then implies that the total Pareto frontier has slopes $V_+'(u) = \{\max\{-c_i, -k\}\}_{i = 1}^n$, each (except for $-k$) over a region of length $u_i^{RFI}$. At the slope of $-k$, either tool can be used. 
Lemma \ref{l: slope of the pareto frontier} also implies the stopping times $T_i^*(h)$, if they exist, are nested in the sense that $T_i^*(h) \leq T_j^*(h)$ if $c_i, c_j < k$ and $c_i < c_j$ (since we provide incentives along cheaper dimensions first). 

This is illustrated below for the case $n = 3$, $k = 1$, $c_i \in \{\frac12, \frac34, \frac54\}$. 

\begin{figure}[h]
	\centering
	\begin{tikzpicture}[scale=0.9]
		\draw[black] (0,0) -- (0,7);
		\draw[black] (-0.8,0.8) -- (11,0.8);

		\node at (0,7) [anchor=south] {\textit{$V(u, \mu_0, \delta)$}};

		\node at (11,0.8) [anchor=west, black] {\textit{$u$}};
		
		\draw[dodgerblue] (0.2,6) -- (4, 5);
		
        \draw[burgundy] (4,5) -- (6, 4);

        \draw[chocolate] (6, 4) -- (10, 0);

        \draw[decorate,decoration={brace,amplitude=10pt}, thin] (4,-0.25) -- (0, -0.25) node [midway, yshift=-0.7cm]  {\textcolor{dodgerblue}{$c_i = 0.5$}};

        \draw[decorate,decoration={brace,amplitude=10pt}, thin](4,-0.25) -- (6,-0.25) node [midway,yshift=-0.7cm]  {\textcolor{burgundy}{$c_i = 0.75$}};
        
		\draw[decorate,decoration={brace,amplitude=10pt}, thin](11,-0.25) -- (6,-0.25) node [midway,yshift=-0.7cm]  {\textcolor{chocolate}{$-k$ region}};
		
	\end{tikzpicture}

	\caption{Pareto Frontier}
	\label{fig: example}	
\end{figure}
The third step is to show that only $\rho^{BP}$, $\rho^{FI}$ are induced on the path of play less a single time in each dimension. The proof strategy is similar to the squeezing lemma. 

\begin{lemma}[Squeezing Lemma, Redux]
\label{l: squeezing redux}
    There exists an optimal contract $\sigma$ where for $\bb{P}^\sigma$ almost all histories $h^\xi$, 
    \[ \#\{t : \rho_i(h^\xi) \not\in \{\rho^{BP}, \rho^{FI}\}\} \leq 1. \]
\end{lemma}
\begin{proof}
Suppose not. Define 
\begin{align*}
    & \mcal G_i(h, \sigma) = \#\{t : \rho(h^\xi; \sigma) \not\in \{\rho^{BP}, \rho^{FI}\} \} \text{  and  } \mcal G_i(\sigma) = \{h: \mcal G_i(h, \sigma) \geq 2\}.
\\ \text{ and also let } & \tilde T_i^*(h; \sigma) = \inf\{t: \rho_i(h^\xi; \sigma) \not\in \{\rho^{BP}, \rho^{FI}\}\} 
\end{align*}
be the first time at which $\rho_i(h^\xi; \sigma)$ is neither full information or Bayesian persuasion. 

Now fix some on-path history $h \in \mcal G_i(\sigma) \cap \bb{P}^\sigma$, so that for $\xi = \tilde T_i^*(h; \sigma)$, one has $\rho_i(h^\phi; \sigma) \not\in \{\rho^{BP}, \rho^{FI}\}$ for $\phi = \xi, \zeta$ where $\zeta \geq \xi$, $h^\zeta \succ h^\xi$. Suppose $\supp(\rho(h^\xi; \sigma)) = \{0, \mu_i(h^\xi)\}$ and $\supp(\rho(h^\zeta; \sigma)) = \{0, \mu_i(h^\zeta)\}$. Consider now small $\varepsilon < \mu_i(h^\xi) - \frac12$ and experiments $\tilde \rho_i(h^\xi; \sigma)$ and $\tilde \rho_i(h^\zeta; \sigma)$ such that $\supp(\tilde \rho_i(h^\xi; \sigma))) = \{0, \mu_i(h^\xi) - \varepsilon\}$ and $\supp(\tilde \rho_i(h^\zeta; \sigma)) = \{0, \mu_i(h^\zeta) + \eta(\varepsilon)\}$ for some $\eta(\varepsilon)$ vanishing as $\varepsilon \to 0$. By Lemma \ref{l: example optimal}, both of these changes are experiments that remain on the Pareto frontier. By Lemma \ref{l: slope of the pareto frontier}, there is moreover some $\eta(\varepsilon)$ such that by changing $\rho_i(h^\phi; \sigma) \to \tilde \rho_i(h^\phi; \sigma)$ for $\phi \in \{\zeta, \xi\}$ and identifying the utility promises after $\mu_i$ to the newly induced beliefs (i.e. changing nothing else), one remains on the Pareto frontier and in particular promises the same amount of utility to the agent at all histories. 

Thus, by logic similar to that in Lemma \ref{l: squeezing lemma} (exploiting the fact the Pareto frontier is linear with slope equal to $\max\{-c_i, -k\}$ at histories $h^{\tilde T_i^*(h; \sigma)}$), it is possible to modify $\sigma$ to some optimum $\tilde \sigma$ such that either $\rho(h^\xi; \tilde \sigma) = \rho^{BP}$ or $h \not\in \mcal G_i(\sigma)$. 
\end{proof}

Now take an optimum satisfying Lemma \ref{l: squeezing redux} and apply the process in Theorem \ref{t: transfers as last resort} so transfers are a last resort; call the result $\sigma^*$, noting $\mcal G_i(\sigma)$ is empty. Define the time 
\[ T_i^*(h) = \inf\{t : \rho_i(h; \sigma^*) = \rho^{FI}\} - 1. \]
Then clearly this stopping time satisfies the condition; we are done if we can show that $T_i^*(h) < \infty$ for $\bb{P}^{\sigma^*}$ almost all histories. 

Suppose not. This implies that $\rho_i = \rho^{BP}$ always. If $c_i < k$ for all $i$, then this is clearly optimal. Otherwise, if there exists $c_i > k$, then by Proposition \ref{p: feasibly optimal actions} it must be that $\alpha_i(h^\xi)(\mu) = 1$ for all $h^\xi$ and $\mu$. But for this to be obedient at $\mu_i = \delta_0$ either the agent's continuation utility must be greater than $0$ following this realization or transfers must occur. Since transfers are a last resort and $\rho_i = \rho^{BP}$ always, transfers cannot motivate the agent. Since $\rho_i = \rho^{BP}$ always, the agents' payoff in each dimension is equal to their no-information outside option baseline. Hence it is impossible for this to be obedient, a contradiction. Thus transfers are used along almost every history. 
\end{proof}

\newpage 
\section*{SUPPLEMENTAL APPENDIX}
\appendix
\section*{Appendix B: The Strong Topology on Contracts}

\makeatletter\def\@currentlabel{Appendix B.}\makeatother
\label{Appendix B}

\setcounter{lemma}{0}
\renewcommand{\thelemma}{B.\arabic{lemma}}

\setcounter{proposition}{0}
\renewcommand{\theproposition}{B.\arabic{proposition}}

\setcounter{definition}{0}
\renewcommand{\thedefinition}{B.\arabic{definition}}

\setcounter{corollary}{0}
\renewcommand{\thecorollary}{B.\arabic{corollary}}

In this appendix I formally define the \emph{strong topology} for contracts $\sigma: H^\infty \to \mcal X^\infty$ where $\mcal X$ is a compact subset of a normed topological space\footnote{Note by the Kuratowski embedding theorem assuming the existence of the norm is without loss of generality, because every metric space can be isometrically embedded in a normed space.}, $H$ is some finite alphabet\footnote{I remark here that the space of histories in the paper is not drawn from some finite alphabet since $\Delta(\Theta) \times A$ is uncountable.
To apply the strong topology, apply the following transformation. First, reduce $\Sigma$ to equivalence classes modulo $\bb{Q}^\sigma$, $\Sigma/\bb{Q}^\sigma$: that is, $\sigma \equiv \sigma'$ if $\bb{Q}^\sigma = \bb{Q}^{\sigma'}$. Second, use the revelation principle to write the space of contracts $\tilde \Sigma$ where $S = A$ directly; this is without loss of generality up to $\bb{Q}^{\sigma}$. Thus $\tilde \Sigma/\bb{Q}^{\sigma} = \Sigma/\bb{Q}^{\sigma}$, and we can endow the latter with the strong topology as we can endow the former with the strong topology. Thus the strong topology can be invoked in the proof of Theorem \ref{t: transfers as last resort} on the space of equivalence classes up to $\Sigma/\bb{Q}^{\sigma}$, which is sufficient for our purposes to apply the squeezing lemma and conclude that transfers are a last resort.} and $H^\infty$ and $\mcal X^\infty$ denote the product topological spaces over $H$ and $\mcal X$, respectively. 
It will be useful to identify contracts $\sigma$ with their \emph{basis decompositions}, i.e. write it as a sequence of outputs of time-$\xi$ histories where the $t$-th coordinate is the element of $\mcal X$ mapped by the time-$\xi$ projection of histories, i.e. 
\[ \sigma(h) = \{\pi_\xi\sigma(h)\}_{\xi = 0}^\infty = \{\sigma(h^\xi)\}_{\xi = 0}^\infty \]
where $h^\xi$ is the projection of $h \in H^\infty$ onto its time-$\{\zeta \leq \xi\}$ cylinder and $\sigma(h^\xi): h^\xi \to \mcal X$ maps the time-$\xi$ projection onto an individual element of $\mcal X$. 

To give an economic interpretation, say $H^\infty$ is the set of infinite histories, $h^\xi$ is the set of time-$\xi$ histories, $H$ is the alphabet of public signals, and $\mcal X$ is the (potentially mixed) action set. $\sigma$ is then a strategy in the corresponding repeated game. 
This is the context in which these objects are used in this paper, and a version of this topology appears in \cite{luo2024marginal}'s Lemma 3. In this appendix, I formally prove and collect several relevant definitions and facts about this topological space. 
Throughout, endow $H$ with the discrete metric and metrize $\mcal X$ with its norm denoted by  $||\cdot||_\infty$. 

\begin{definition}
    Say that $\sigma_n \to \sigma$ in the \emph{strong topology} if and only if 
    \[ \lims_{n \to \infty} d(\sigma_n, \sigma) =  \lims_{n \to \infty} \sup_{h \in H^\infty} \sum_{\xi = 0}^\infty \frac{1}{2^\xi} ||\sigma_n(h^\xi) - \sigma(h^\xi)||_\infty = 0. \]
\end{definition}

\begin{proposition}
\label{p: contract space compact}
    Let $\Sigma$ be the set of all strategies. $\Sigma$ is compact in the strong topology. 
\end{proposition}
\begin{proof}
The goal is to adapt a sufficiently general version of the functional Arzela-Ascoli theorem to prove that $(\Sigma, d)$ is sequentially compact. 
First, the space $\Sigma$ is equicontinuous; note that 
\[ d(\sigma(h), \sigma(h')) = \sum_{\xi = 0}^\infty \frac{1}{2^\xi} ||\sigma(h^\xi) - \sigma(h^{'\xi})||_\infty \]
and so in particular if any two histories $h, h'$ agree for the first $T$ periods (in which case the distance $d(h, h')$ is at most $\frac{1}{2^{T - 1}}$) then it must be that the distance of their images is separated by at most $\frac{1}{2^{T - 1}}$. Since this is true for any $\sigma$ and at any history, the modulus of continuity for functions in $\Sigma$ is independent both of the choice of $h$ and of the choice of $\sigma$ and hence $\Sigma$ is equicontinuous. Moreover, this space of contracts is closed; for any sequence $\{\sigma_n\}$ which converges to some $\sigma$, $\sigma$ is itself a function defined on the cylinders as  
\[ \sigma(h^\xi) = \lims_{n \to \infty} \sigma_n(h^\xi)  \]
where convergence of $\sigma_n(h^\xi)$ is guaranteed as $\mcal X$ is compact. 
Finally, for any fixed $h \in H^\infty$, the closure of the set $\{\sigma(h)\}_{\sigma \in \Sigma}$ is a closed subset of a compact set (namely, $\mcal X^\infty$) and hence compact. Note also $H^\infty$ is compact as well. This implies we satisfy Theorem 47.1 of \cite{munkres2000} and hence $\Sigma$ is sequentially compact (though it is proven in the topology of compact convergence---this coincides with our topology when the ambient space of the domain is compact). 
\end{proof}

Now suppose $\mcal X = \Delta(H)$, so that each time-$\xi$ history is mapped to a (random) new letter of the alphabet. Then for any any function $\sigma$, note that there is a natural measure defined on $\mcal H^\infty$ generated by the cylinders $\bb{P}_\xi^{\sigma}(h^\xi) = \bb{P}_{\xi - 1}^\sigma(h^{\xi-1})\sigma(h^{\xi-1})(x)$, where $h^\xi = \{h^{\xi-1}, x\}$ for $x \in H$. Let the Kolmogorov extension of the consistent family $\{P_\xi^\sigma\}$ to the cylinder $\sigma$-algebra be $\bb{P}^\sigma$. 
I conclude with the following result. 

 \begin{proposition}
     If $\sigma_n \to \sigma$ in the strong topology, $\bb{P}^{\sigma_n} \to \bb{P}^\sigma$ pointwise (i.e. on cylinder sets). 
 \end{proposition}
 \begin{proof}
     We first show $\bb{P}_\xi^{\sigma_n} \to \bb{P}_\xi^{\sigma}$. Clearly this is true when $\xi = 0$, since the starting measure is trivial. Inductively, suppose $\bb{P}_{\xi - 1}^{\sigma_n} \to \bb{P}_{\xi - 1}^\sigma$ for every history $h$. Then 
     \[ \bb{P}_{\xi}^{\sigma_n}(h) = \bb{P}_\xi^{\sigma_n}(h^\xi) = \bb{P}_{\xi - 1}^{\sigma_n}(h^{\xi-1})\sigma_n(h^{\xi-1})(x); \] 
     since the inductive hypothesis implies $\bb{P}_{\xi - 1}^{\sigma_n}(h^{\xi - 1}) \to \bb{P}_{\xi - 1}^{\sigma}(h^{\xi - 1})$ and $\sigma_n \to \sigma$ in the strong topology, one has 
     \[ \bb{P}_{\xi - 1}^\sigma(h^{\xi-1})\sigma(h^{\xi-1})(x) = \bb{P}_\xi^\sigma(h^\xi) \]
as desired. Taking $\xi \to \infty$ then gives the result. 
 \end{proof}

\appendix
\section*{Appendix C: Incentivizability with Continuous Actions}

\makeatletter\def\@currentlabel{Appendix C.}\makeatother
\label{Appendix C}

\setcounter{lemma}{0}
\renewcommand{\thelemma}{C.\arabic{lemma}}

\setcounter{proposition}{0}
\renewcommand{\theproposition}{C.\arabic{proposition}}

\setcounter{definition}{0}
\renewcommand{\thedefinition}{C.\arabic{definition}}

\setcounter{corollary}{0}
\renewcommand{\thecorollary}{C.\arabic{corollary}}

Suppose now $\Theta$ was some compact set and $A$ was a compact convex subset of $\bb{R}$. To match the assumptions of Section 4 and to avoid the mathematics of continuum space Markov kernels, assume the state is drawn i.i.d. across periods from some Borel measure $\mu_0 \in \Delta(\Theta)$, fully supported in $\Theta$. Moreover, suppose the following are true on Sender and Receiver payoffs $v(a, \theta)$ and $u(a, \theta)$. 

\begin{assumption}
\label{ass: interiority assumption}
$u(a, \theta)$ and $v(a, \theta)$ are $\mcal C^2$ and strictly concave in $a$ for each fixed $\theta$ and jointly continuous in $(a, \theta)$. For all $\mu$ there exists $a^*(\mu) \in A^o$ such that $\frac{\partial}{\partial a} \bb{E}_\mu[u(a^*(\mu), \theta)] = 0$. 
\end{assumption}

The first part of the assumption guarantees that a first order approach is sufficient for each fixed belief $\mu$. The second part guarantees that there is an interior solution to the first order condition. Assumption \ref{ass: interiority assumption} is thus a regularity condition that guarantees that the agents' maximization problem is well-defined and interior regardless of the induced belief. It is satisfied by leading examples of communication, including the quadratic loss setting of \cite{CrawfordSobel1982}. 

Why work with a continuum? In general the key condition for incentivizability is that there is some scope for local incentives to be used to perturb Receiver's optimal action at some belief. In general (in the static problem), Receiver must be paid by $k(\bb{E}_\mu[u(a, \theta) - u(a', \theta)])$ to take action $a$ instead of $a'$; this gives Sender a benefit of $\bb{E}_\mu[v(a, \theta) - v(a', \theta)]$. When actions are discrete, there is no guarantee that the benefit at any induced belief to Sender is greater than the cost of inducing any action. However, on the continuum, local changes to Receiver's payoff at their optimum are second order, while the benefit to changing an action to Sender is first order. Thus, so long as Sender and Receiver's favorite actions do not coincide, one has that 
\[ \lims_{a \to a^*(\mu_0)} \left|\frac{\bb{E}_\mu[v(a, \theta) - v(a^*(\mu_0), \theta)]}{a - a^*(\mu_0)}\right| > \lims_{a \to a^*(\mu_0)} k\left|\frac{\bb{E}_\mu[u(a, \theta) - u(a^*(\mu_0), \theta)]}{a - a^*(\mu_0)}\right| \]
and thus Sender will find it profitable to change Receiver's action locally, as
\[ \lims_{a \to a^*(\mu)} \left| \frac{\bb{E}_{\mu}[v(a, \theta) - v(a^*(\mu), \theta)]}{\bb{E}_{\mu}[u(a, \theta) - u(a^*(\mu), \theta)]} \right| > k \]
and so an argument similar to the one which undergirds Proposition \ref{p: feasibly optimal actions} applies. Hence one would expect that primitives are in general incentivizable on the continuum, since they satisfy the condition of Proposition \ref{p: wrr if static k-cavification}. 

The main result of this appendix is thus Proposition \ref{p: continuous incentivizability}, stated below. Say Sender \emph{does not attain first best} at $\delta = 0$ if 
\[ \text{cav}(V_0)(\mu_0) \neq \bb{E}_{\mu_0}\left[\max_{a \in A} v(a, \theta) \right] \]
that is, Sender does not attain their highest possible payoff. 
Note if this is the case then Sender attains their highest possible payoff in the dynamic problem as well for any $\delta$, since the repeated static optimal contract is a feasible dynamic contract. So long as this is not the case, primitives are incentivizable with a continuum of actions. 

\begin{proposition}
    \label{p: continuous incentivizability}
    Suppose Sender does not attain first best at $\delta = 0$. Then primitives $\{u, v, k, \delta\}$ are incentivizable. 
\end{proposition}
\begin{proof}
    By Proposition \ref{p: wrr if static k-cavification}, it is sufficient to show $\mcal V^*(\mu_0, 0) > \text{cav}(V_0)(\mu_0)$. Suppose not. Then the static optimal solution without transfers is also the solution to the problem with transfers at $\delta = 0$. Let $\{\rho, t, u', \alpha\} \in \mcal F(0, \mu_0)$ be one such solution to the problem where $u' \equiv t \equiv 0$, so that $\alpha(\mu) = a^*(\mu, \mathbf{0})$ for all $\mu \in \supp(\rho)$. 
    Because Sender does not do first best, there exists a set $\mcal C$ with $\rho(\mcal C) > 0$ such that $\frac{\partial}{\partial a} \bb{E}_\mu[v(a^*(\mu, \mathbf{0}), \theta)] \neq 0$ for all $\mu \in \mcal C$. Without loss of generality, suppose this term is positive for all $\mu \in \mcal C$. Define the following actions on $\mcal C$: 
    \[ \tilde a(\mu) = a^*(\mu, \mathbf{0}) + \varepsilon(\mu) \]
    where $\varepsilon(\mu) > 0$ is chosen so that 
    \[ \left|\frac{\bb{E}_\mu[v(a^*(\mu, \mathbf{0}) + \varepsilon(\mu), \theta) - v(a^*(\mu, \mathbf{0}), \theta)  ]}{\bb{E}_\mu[u(a^*(\mu, \mathbf{0}) + \varepsilon(\mu), \theta) - u(a^*(\mu, \mathbf{0}), \theta)]} \right| > k \]
    noting such an $\varepsilon(\cdot)$ exists for each $\mu$ because $\frac{\partial}{\partial a} \bb{E}_\mu[u(a^*(\mu, \mathbf{0}), \theta)] = 0$ but $\frac{\partial}{\partial a} \bb{E}_\mu[v(a^*(\mu, \mathbf{0}), \theta)] > 0$. 
    On $\mcal C^c$, do not change the action. 
    From here, set 
    \[ \tilde t(\mu) = \bb{E}_\mu[u(a^*(\mu, \mathbf{0}), \theta) - u(a^*(\mu, \mathbf{0}) + \varepsilon(\mu), \theta)]\]
    on $\mcal C$, and $0$ otherwise. The tuple $\{\rho, \tilde t, 0, \tilde a\}$ is then a feasible tuple, and moreover by construction gives Sender a strictly higher payoff on $\mcal C$ and the same payoff on $\mcal C^c$. Thus, it is a profitable deviation, contradicting our assumption $\{\rho, 0, 0, \alpha\}$ was optimal. 
\end{proof}

The continuum is not without its own challenges. While Proposition \ref{p: weak ray regularity} and \ref{p: wrr if static k-cavification} lift immediately for any action space (even those violating Assumption \ref{ass: interiority assumption}), Proposition \ref{p: contract space compact} does not, and hence it is not immediately clear that Theorem \ref{t: transfers as last resort} need to hold. Intuitively, this is because the entire space of contracts $\Sigma$ need not be compact absent topological restrictions on its codomain\footnote{The cardinality of the codomain is too large to guarantee equicontinuity on the space of functions, so one cannot apply functional Arzela-Ascoli.} However, for an arbitrarily fine grid approximation of the continuum, a continuity argument implies that primitives are both incentivizable and I can guarantee that transfers are a last resort. 

Proposition \ref{p: continuous incentivizability}, combined with Proposition \ref{p: weak ray regularity}, implies the following corollary, which shows that in general on the continuum, transfers are strictly backloaded. 

\begin{corollary}
    Suppose Assumption \ref{ass: interiority assumption} holds for $(u, v)$. Then for any $k > V_+'(0)$ and $\delta < 1$, if Sender does not attain first best when $\delta = 0$, then transfers are strictly backloaded. 
\end{corollary}

\appendix
\section*{Appendix D: The Static K-Cavification: An Example}

\makeatletter\def\@currentlabel{Appendix D.}\makeatother
\label{Appendix D}

\setcounter{lemma}{0}
\renewcommand{\thelemma}{D.\arabic{lemma}}

\setcounter{proposition}{0}
\renewcommand{\theproposition}{D.\arabic{proposition}}

\setcounter{definition}{0}
\renewcommand{\thedefinition}{D.\arabic{definition}}

\setcounter{assumption}{0}
\renewcommand{\theassumption}{D.\arabic{assumption}}

\setcounter{corollary}{0}
\renewcommand{\thecorollary}{D.\arabic{corollary}}

I give an example to highlight why the $\mcal K$-cavification substantially simplifies the problem of finding the concavified transfer-augmented value function. 

Consider the interaction between a manager and an employee who must choose between two differentiated projects to pursue. The two projects are differentiated horizontally by the employees' fit for the project ($\theta \in \{\theta_0, \theta_1\}$)\footnote{For example, how much the employee would enjoy being on that project, how complementary their skill sets are to the objective, the difficulty of the project, etc.}, which the manager privately observes. 
The manager has a (known) preferred project $a_1$ that they would always like the employee to work on, regardless of the characteristics of the other project $a_0$\footnote{This can be, for example, because one of the two projects matters for the manager's promotion package while the other does not.}. However, the employee would like to work on whichever project is a better fit for them (e.g. $a_0$ at $\theta_0$ and $a_1$ at $\theta_1$), though ex-ante they cannot observe the project's fit (for example, because the manager can conceal information about each project before they decide which one to join). The employee also always has an outside option, $a_2$ (e.g. transferring to a different project), which guarantees some safe payoff, normalized to $0$, which is the worst option for the manager. 
An example of payoffs capturing the above scenario are given below. Fix $k = 1$ throughout.
\begin{table}[htb]
\centering
\begin{tabular}{|c|c|c|c|}
\hline
S/R & $a_0$ & $a_1$ &  $a_2$ \\ \hline
$\theta_0$ & $0,1$ & $2.5,-2$ & $-0.5, 0$ \\ \hline
$\theta_1$ & $0,-2$  & $2.5,1$ & $-0.5, 0$ \\ \hline
\end{tabular}
\end{table}

Before the interaction, the worker is biased \textit{against}  the manager---they believe the state is likely to be $\theta_0$ with prior belief  $\mu_0 = \mu_0(\theta = \theta_1) = \frac16$. The manager's indirect value function takes the form 
\[ V(\mu) = \begin{cases} 0 \text{  for  } 
\mu \in [0, \frac13] \\ 
- \frac12 \text{  for  } \mu \in (\frac13, \frac23) \\ 
\frac52 \text{  for  } \mu \in [\frac23, 1]
\end{cases} \]
The extremal beliefs are then simply $\{0, \frac13, \frac23, 1\}$. At $0$, it is optimal to pay nothing, as is the case at $1$, and let the worker act optimally. A computation shows it is optimal to pay to induce action $a_1$ at $\frac13$, but again pay nothing at $\frac23$. These four simple computations are enough to globally compute the value of transfers and persuasion, both at $\mu_0 = \frac16$ and at any arbitrary prior belief. The result is graphed below. 

\begin{figure}[ht]
\centering
   \begin{tikzpicture}[scale=1.1]
    \begin{axis}[
        axis lines=middle,
        xtick={0,0.1667, 0.333, 0.667,1}, 
        xticklabels={$0$, $\mu_0$, $\frac{1}{3}$, $\frac{2}{3}$, $1$},
        ytick={-1/2,0,2.5},
        xlabel={$\mu$},
        ylabel={$V(\mu)$},
        domain=0:1,
        samples = 100
    ]
        \addplot[
            color=dodgerblue,
            thick
        ] coordinates {
            (0,0) (1/3,0)
        };
        \addplot[
            color=dodgerblue,
            thick
        ] coordinates {
            (1/3,-0.5) (2/3,-0.5)
        };
        \addplot[
            color=dodgerblue,
            thick
        ] coordinates {
            (2/3,2.5) (1,2.5)
        };

        \addplot[only marks, mark=o, mark size=3pt, dodgerblue] coordinates {(1/3,-0.5) (2/3,-0.5)};
        \addplot[only marks, mark=*, mark size=3pt, dodgerblue] coordinates {(1/3,0) (2/3,2.5)};

        \addplot[
            color=chocolate,
            thick,
            dashed
        ] coordinates {
            (0,0) (2/3,2.5)
        };

        \addplot[
            color=burgundy,
            thick,
        ] coordinates {
            (0,0) (1/3,1.5)
        };

        \addplot[
            color=burgundy,
            thick,
        ] coordinates {
            (1/3,1.5) (2/3,2.5)
        };

         \addplot[
            color=burgundy,
            thick,
        ] coordinates {
            (2/3,2.5) (1, 2.5)
        };

        \addplot[only marks, mark=*, mark size=3pt, burgundy] coordinates {(1/3,1.5)};
        
        \addplot[
                    color=black,
                    thin,
                    dashed
                ] coordinates {
                    (1/6,0) (1/6,3/4)
                };
        \node[draw=black, fill=white, font=\scriptsize, anchor=west] at (axis cs: 0.45, 0.75) {
            \begin{tabular}{l}
                \textcolor{dodgerblue}{Value Function} \\
                \textcolor{chocolate}{Persuasion Only} \\
                \textcolor{burgundy}{Persuasion + Transfers}
            \end{tabular}
        };
  \end{axis}
\end{tikzpicture}
\caption{Value with Transfers}
\label{fig: example part 2}
\end{figure}
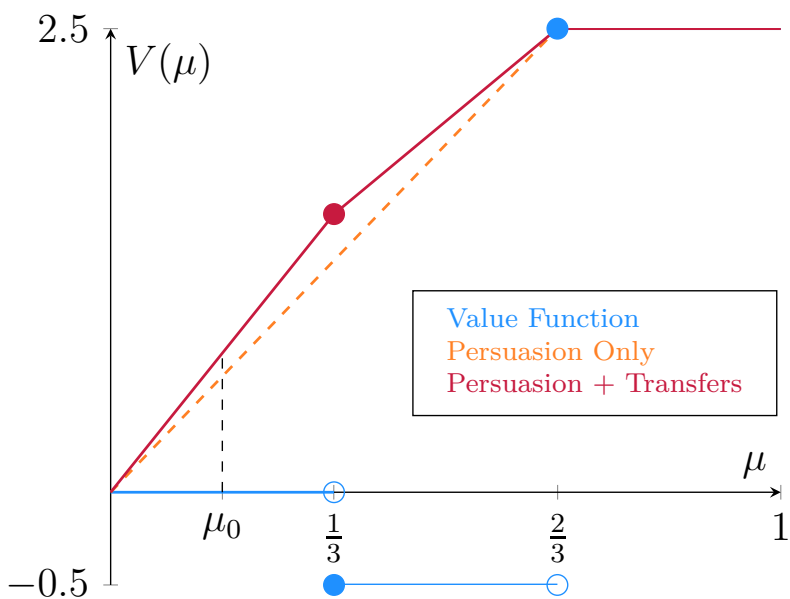

Given the $\mcal K$-cavification procedure, it is now easy to see that the manager values persuasion (and transfers) jointly at $[0, \frac23)$, and their joint value outperforms their value from only persuasion on that same interval. Absent computing the extremal beliefs, this would have required first computing $V^t(\mu)$ belief-by-belief and then concavifying the resulting (potentially complicated) function. 

Suppose there is an equilibrium where payments and persuasion are both used (as at $\mu_0 = \frac16$ in the example). As $\delta \to 0$, upper hemi-continuity of the best response then implies that there cannot be optima for arbitrarily small $\delta$ where transfers are used only at histories where the state is fully revealed. Why does this not contradict Theorem \ref{t: transfers as last resort}? Note that in the statement of Theorem \ref{t: transfers as last resort} there is possibly one period of indeterminancy where transfers are used but the state is not yet fully revealed: full revelation need to occur only after transfers are positive at \textit{successor} histories. In the static case, that period of indeterminancy is exactly the first period.

\appendix
\section*{Appendix E: Proof of Proposition \ref{p: FE properties}}

\makeatletter\def\@currentlabel{Appendix E.}\makeatother
\label{p: FE properties proof}

\setcounter{lemma}{0}
\renewcommand{\thelemma}{F.\arabic{lemma}}

\setcounter{proposition}{0}
\renewcommand{\theproposition}{F.\arabic{proposition}}

\setcounter{definition}{0}
\renewcommand{\thedefinition}{F.\arabic{definition}}

\begin{proof}
Fix $\delta \in (0, 1)$. Define an operator $\mcal O: \mcal B([-C, C] \times \Delta(\Theta)) \to  \mcal B([-C, C] \times \Delta(\Theta))$, where $\mcal B(X)$ is the set of bounded functionals from $X \to \bb{R}$. 
    Let $\mcal O$ be defined by
 \begin{align*}
\label{equation_1}
\mcal O(V)(\bar u, \mu_0, \delta) = \max_{\{\rho, t, u', \alpha\}} 
   \Big\{ \bb{E}_\rho\Big[\bb{E}_{\mu, \alpha}[(1 - \delta)\,[v(a, \theta) & - k t(\mu, a)] 
   + \delta V(u'(\mu, a), M\mu, \delta)]\Big] \Big\}
\tag{FE} \\[1em]
\text{s.t.}\quad 
\bb{E}_{\mu}[(1 - \delta) [u(a, \theta) + t(\mu, a)] + \delta u'(\mu, a)] 
&\geq \bb{E}_\mu[(1 - \delta) u(a', \theta) + \delta \und U(M\mu)]
\tag{IC} \\
&\hspace{0.5em}\forall \mu \in \supp(\rho),\; a \in \supp(\alpha(\mu)),\; a' \in A \\[1em]
\bb{E}_\rho[\bb{E}_{\mu, \alpha}[(1 - \delta)[u(a, \theta) + t(\mu, a)] + \delta u'(\mu, a)]] 
&\geq \bar u 
\tag{PK} \\[1em]
\bb{E}_\rho[\mu] 
&= \mu_0 
\tag{BP} \\[1em]
\max\{(1 - \delta) t(\mu, a),\, |u'(\mu, a)|\} 
\in [0, C] \quad \forall  & \mu \in \supp(\rho), a \in \supp(\alpha(\mu)), 
\tag{BD}
\end{align*}
Take any $W \geq V$; clearly monotonicity is satisfied. Similarly, any constant $\beta < 1$ can be pulled out of the maximum without affecting the constraints and hence discounting is also satisfied. Blackwell's sufficient conditions for a contraction mapping then imply that there is a unique fixed point of $\mcal O$, which is exactly the value function $V$. 
Moreover, $V$ is continuous, as the mapping $\mcal O$ restricted to the (closed) space of continuous functions maps into itself, because the objective and constraints vary continuously in $\bar u, \mu_0$. 

That $V$ is nonincreasing in $\bar u$ follows immediately from the fact that the set of feasible solutions is weakly decreasing in $\bar u$ (it can only strictly tighten the promise keeping constraint). 
Concavity follows by noting that $\mcal O(V)(u, \mu_0, \delta)$ is concave in $u$ whenever $V$ is concave since the convex combination of a feasible solution at $u$ and $u'$ is feasible at $\alpha u + (1 - \alpha)u'$ (as the constraints are linear in the relevant arguments). Hence, $\mcal O$ maps the set of concave functions to itself, and because the space of concave functions is a closed subset (in the supremum norm) of all functions, one can again apply Banach's fixed point theorem as before. 

Finally, the bound on its right derivative. Fix any $\mu$, $u$, and $\varepsilon > 0$, and let $\{\rho, t, u', \alpha\} \in \mcal F(u, \mu)$. Note $\{\rho, t + \frac{\varepsilon}{1 - \delta}, u', \alpha\}$ is feasible at $(u + \varepsilon, \mu)$ and gives Sender a payoff of $V(u, \mu) - k\varepsilon$ exactly. Thus, $V(u + \varepsilon, \mu) > V(u, \mu) - k\varepsilon$. Then 
\[ \lims_{\varepsilon \to 0_+} \frac{V(u + \varepsilon, \mu) - V(u, \mu)}{\varepsilon} \geq  \lims_{\varepsilon \to 0_+} \frac{V(u, \mu) - k\varepsilon - V(u, \mu)}{\varepsilon} =  \lims_{\varepsilon \to 0_+} -\frac{k\varepsilon}{\varepsilon} = -k\]
as desired. This finishes the proof. 
\end{proof}

\end{document}